\documentclass[twocolumn]{aastex62}

\usepackage{tabularx}
\usepackage{enumitem}
\usepackage{amsmath}

\newcolumntype{S}{>{\centering\arraybackslash}X}

\accepted{2018 March 23 to PASP}

\shorttitle{RIT Polarization Imaging Camera}
\shortauthors{Vorobiev et al.}

\begin{document}

\title{Astronomical Polarimetry with the RIT Polarization Imaging Camera}

\correspondingauthor{Dmitry Vorobiev}
\email{vorobiev@cis.rit.edu}

\author{Dmitry V. Vorobiev}
\author{Zoran Ninkov}
\affil{Center for Imaging Science, Rochester Institute of Technology, 54 Lomb Memorial Drive, Rochester, NY 14623, USA}
\author{Neal Brock}
\affil{4D Technology Corporation, 3280 E. Hemisphere Loop, Suite 14, Tucson, AZ 85706, United States}



\begin{abstract}
{\small In the last decade, imaging polarimeters based on micropolarizer arrays have been developed for use in terrestrial remote sensing and metrology applications. Micropolarizer-based sensors are dramatically smaller and more mechanically robust than other polarimeters with similar spectral response and snapshot capability. To determine the suitability of these new polarimeters for astronomical applications, we developed the RIT Polarization Imaging Camera to investigate the performance of these devices, with a special attention to the low signal-to-noise regime. We characterized the device performance in the lab, by determining the relative throughput, efficiency, and orientation of every pixel, as a function of wavelength. Using the resulting pixel response model, we developed demodulation procedures for aperture photometry and imaging polarimetry observing modes. We found that, using the current calibration, RITPIC is capable of detecting polarization signals as small as $\sim0.3\%$. The relative ease of data collection, calibration, and analysis provided by these sensors suggest than they may become an important tool for a number of astronomical targets.}
\end{abstract}




\section{Introduction} \label{sec:introduction}
Measurements of light polarization are responsible for discoveries in nearly all areas of astronomy. Starting with the pioneering observations by Arago in 1811 \citep{Arago1855}, polarization has been measured for most astronomical objects: from the Sun \citep{Hale1908} and Solar System planets \citep{Lyot1929, Coffeen1969, Gehrels1969, Hansen1974}, comets \citep{Hines2016}, and asteroids \citep{Dollfus1971,Belskaya2015}, to debris disks and planetary systems around other stars \citep{Kolokolova2015}, proto-planetary nebulae \citep{Ueta2007}, active galactic nuclei \citep{Antonucci1985}, and the large scale structure of the universe \citep{Planck2016}.

Because polarization is an intrinsic property of light, polarimetry provides a window into light-matter interactions across a wide range of spatial scales and astrophysical phenomena. Polarization analysis has been used to infer the microscopic geometry of scattering/absorbing particles in many different environments (\textit{e.g.} the sulfuric acid droplets in the Venusian atmosphere \citep{Hansen1974b}, dust grains in the interstellar medium \citep{Hiltner1949, Hall1949} or debris disks around young stars \citep{Perrin2014}). On the macroscopic scale, polarimetry provides a means of overcoming the degeneracy associated with observing 2D projections of 3D objects, as was demonstrated through observations of debris disks \citep{Perrin2014} and dust shells around the Cepheid variable RS Puppis \citep{Kervella2014}. Polarimetry can also be used to infer the macroscopic geometry of unresolved objects, like the accretion disk and torus environments around active galactic nuclei \citep{Antonucci1985}. 

Finally, polarimetry is a powerful probe of many astrophysical processes. It is one of the few tools available in the study of magnetic fields. \cite{Hale1908} used a spectropolarimeter to measure the magnetic field of Sun spots, whereas the stellar polarization measurements like those of \cite{Hiltner1949} and \cite{Hall1949} are used to study the galactic magnetic field. More advanced polarimeters like CanariCam \citep{Telesco2003} attempt to disentangle the complex magnetic fields associated with inflows and outflows of matter during star formation. 

\newpage
\subsection{Operating Principles of Polarimeters}
The polarization of light is challenging to measure due to a lack of detectors\footnote{This is true for detectors used in the x-ray to far-infrared regime. Many microwave and radio receivers measure the polarization directly; however, the focus of this work is on polarimeters designed for the former.} with inherent sensitivity to the state of polarization. Thus, polarization measurement must be accomplished through a scheme that modulates the intensity of radiation, based on its polarization state. This intensity modulation can then be measured and demodulated to recover the polarization state. 

Over the years, several different methods to modulate the intensity have been developed. The modulation can occur in the spatial, temporal and spectral domains (see reviews by \cite{Hough2006} and \cite{Snik2014}). Polarimeters exhibit a rich fauna of designs; however, they can be roughly separated into single-beam and dual-beam operating modes. Single-beam polarimeters rely on changing the orientation of a polarizing filter (polymer or wire grid) to make a sequence of measurements over a certain period of time. This approach is especially attractive for imaging polarimetry, because large aperture polarizers are inexpensive, light weight, and easy to use, as opposed to large aperture beam-splitting optics. Polarimetry on the \textit{Hubble Space Telescope} is performed with polarizing filters, notably on NICMOS \citep{Hines2000, Batcheldor2009} and the ACS \citep{Avila2017}. Also, observations of the corona during solar eclipses are often made using a polarizer and a film \citep{Molodenskii2009} or electronic camera \citep{Qu2013}.  

The single-beam approach is limited by systematic errors associated with the division-of-time scheme and limited means of dealing with instrumental polarization (other than careful characterization and calibration). The dual-beam approach allows the acquisition of several intensities at the same time, alleviating some requirements on instrumental and source stability. Most often the beam-splitter is a Wollaston prism. Furthermore, dual-beam polarimeters equipped with retarders (a rotating half-wave plate or an electro-optical modulator) can mitigate instrumental effects, downstream of the retarder, by allowing the same detector (or detector pixel) to measure intensities of orthogonal polarization states (\textit{i.e.}, the amount of 0$^{\circ}$ light and 90$^{\circ}$ light).

Bernard Lyot built on the ideas of Savart to construct a polariscope with exquisite sensitivity \citep{Lyot1929}. By the addition of a tilting quartz plate, Lyot was able to slightly polarize the incident light, to bring weakly polarized objects across the detection limit of a typical Savart polariscope, or vice versa. Polarimeters built between 1940 and 1970 employed many techniques and components to deal with instrumental effects and scintillation, including the use of slowly or rapidly variable retarders and depolarizers. In parallel, advances in photometric tools (such as photomultiplier tubes and Digicon arrays) helped improve polarimetric sensitivity. Today, the most sensitive instruments use some version of the dual-beam dual-difference technique: PlanetPol \citep{Hough2006b}, GPI \citep{Macintosh2006}, ZIMPOL \citep{Roelfsema2010}, the Extreme Polarimeter \citep{Rodenhuis2012}, POLISH2 \citep{Wiktorowicz2015} \textit{et al.}

Recently, several groups have built polarization-sensitive imaging sensors, using micropolarizer arrays \citep{Nordin1999, Brock2011, Vorobiev2016}. These imaging polarimeters build on the advantages offered by other single-beam systems with the addition of snapshot capability. These devices are compact, light, mechanically robust and simple to operate, with minimal power requirements. Until now, these sensors have been used in metrology applications \citep{Brock2011}; however, some are exploring their use for remote sensing.

In this work, we investigate the suitability of these devices for applications in astronomy. To this end, we designed and built the Rochester Institute of Technology Polarization Imaging Camera (RITPIC). Our goal was not to build an instrument with a specific science goal in mind, as is the case for most polarimeters developed by astronomers. Instead, we are interested in the fundamental sensitivity that can be achieved with these devices and the factors which limit performance during on-sky observation. In this paper we describe the design of RITPIC and the operating principles of micropolarizer-based polarimeters, as they pertain to astronomical polarimetry. We describe the data analysis process we developed and discuss the advantages afforded by (and challenges associated with) these unique polarimeters. To demonstrate RITPIC's capability, we present ``first light'' polarimetry of standard stars, solar system planets, and protoplanetary nebula, performed on the 0.9 m SMARTS telescope at Cerro Tololo Inter-American Observatory.  We conclude with a summary of RITPIC's current (and potential) performance and a discussion of applications that appear most suited for these polarimeters.

\newpage
\section{Instrument Design} \label{sec:instrumentDesign}
\begin{figure}[ht]
\centering
\includegraphics[width=\linewidth]{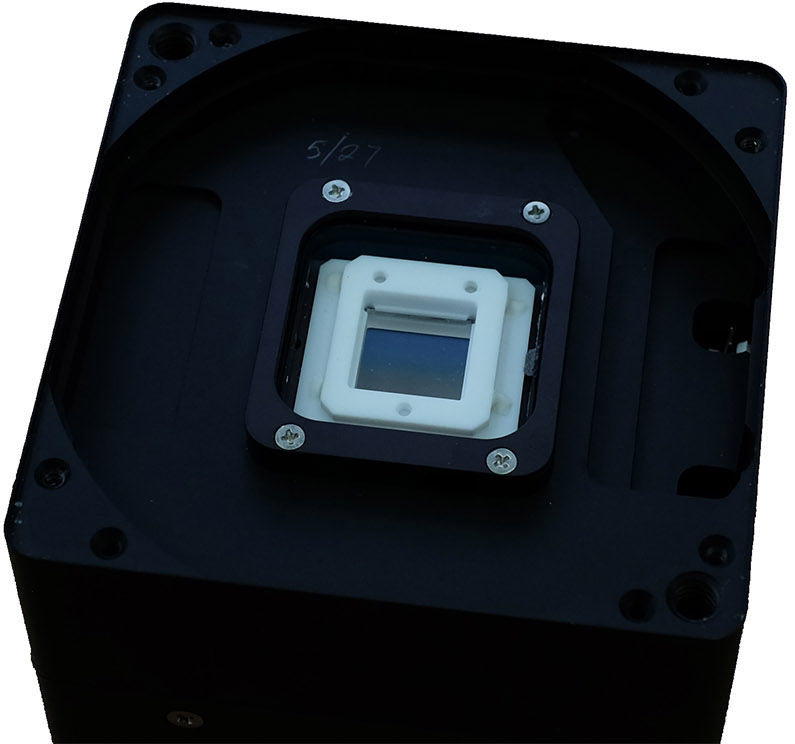}
\caption{RITPIC is built using a KAI-04070 CCD and a MOXTEK, Inc. micropolarizer array. The micropolarizer array (silver square) is aligned and affixed to the CCD using a white ceramic carrier; the alignment was performed by 4D Technology Corp. The polarization-sensitive imager is housed in a Finger Lakes Instrumentation MicroLine camera body, which allows thermoelectric cooling of the sensor. During observations, a baffle is used to cover the white carrier; see Appendix A for more information.}
\label{fig:ritpic}
\end{figure}

\begin{figure}[ht]
\centering
\includegraphics[width=\linewidth]{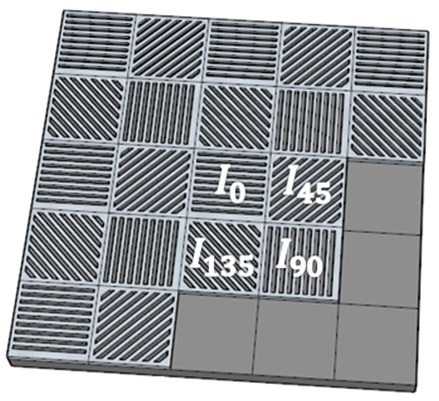}
\caption{RITPIC uses a micropolarizer array to modulate the intensity of the incident light at the focal plane, based on the light's polarization. Each pixel of the CCD is aligned with a single, pixel-sized, wire grid linear polarizer. A set of 4 pixels samples the electric field intensity along the 0$^{\circ}$, 45$^{\circ}$, 90$^{\circ}$, and 135$^{\circ}$ directions.}
\label{fig:mpa_grid}
\end{figure}

The RIT Polarization Imaging Camera (RITPIC) is an imaging polarimeter based on the division-of-focal plane modulation scheme (Figure \ref{fig:ritpic}). RITPIC's detector is a KAI-04070 CCD (see Table \ref{tbl:ccdSpecs} for CCD specifications) in a thermo-electrically cooled camera from Finger Lakes Instrumentation. A micropolarizer array is aligned to the CCD pixel grid, such that each CCD pixel is covered by a pixel-sized wire grid polarizer (Figure \ref{fig:mpa_grid}); the result is an imager with intrinsic sensitivity to the polarization of light. This is analogous to the use of color filter arrays to create color-sensitive imaging arrays.

\begin{table}[ht]
    \centering
    \begin{tabularx}{\linewidth}{ S S }
    \hline
    \hline
         Parameter &\hspace{1.2cm} Value$^a$ \\
         \hline
         Peak QE & \hspace{1.2cm} 52\% at 500 nm  \\
         $\lambda$ Range$^{b}$ &\hspace{1.2cm} 350 - 800 nm  \\
         Charge Capacity  &\hspace{1.2cm}  36,278 $\mathrm{e^-}$   \\
         Read Noise &\hspace{1.2cm} 7.0 $\mathrm{e^-}$ \\
         Gain &\hspace{1.2cm} 0.570 $\mathrm{e^-}$/ADU \\
    \hline
    \end{tabularx}
    \caption{Performance metrics for RITPIC’s detector, the KAI-04070 CCD. Notes:$^a$CCD performance metrics measured using the 1.5 MHz readout mode. $^{b}$Range where QE $>$ 15\%.}
    \label{tbl:ccdSpecs}
\end{table}

The division-of-focal plane (DoFP) modulation technique (see \cite{Tyo2006} and \cite{Snik2014} for overviews of modulation strategies and polarimeter designs) allows the fabrication of extremely small, mechanically robust polarimeters, with no moving parts and a minimal amount of optical and electronic components. The polarization sensor can be based on any detector technology: CCD, active pixel sensor / CMOS, and even advanced hybridized infrared arrays. RITPIC samples the electric field along all necessary orientations in a single exposure (aka ``snapshot'' capability), which is especially useful for targets and scenes with temporal and spatial variability. Nevertheless, the DoFP technique is vulnerable to several sources of error. These errors must be mitigated with proper calibration, observing strategy, and data analysis techniques. 

\section{Performance Characterization} \label{sec:performanceCharacterization}
In this section, we describe the characterization procedure used to determine the transmission, efficiency, and orientation for each pixel in the polarization-sensitive imager. Accurate characterization of these properties is critical to the performance of these devices, because the data acquisition and analysis methods typically used with these sensors do not allow for convenient calibration methods, such as the dual-beam double-difference technique \citep{Donati1990, Semel1993}.

\begin{figure}[ht]
\centering
\includegraphics[width=\linewidth]{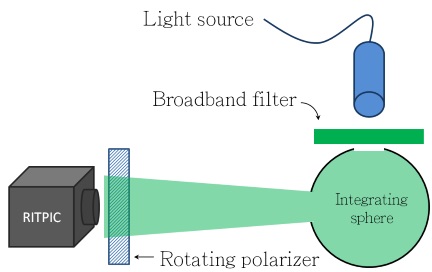}
\caption{The polarization sensors were characterized using an integrating sphere illuminated by a broadband light source and a rotating linear polarizer. Broadband filters were introduced between the light source output and the integrating sphere, to determine the wavelength dependence of the system response.}
\label{fig:labSetup}
\end{figure}

Polarimeters based on micropolarizer arrays have no moving parts and are permanently aligned. Therefore, these systems are extremely stable and do not need frequent calibration. The setup used to characterize the performance of RITPIC is shown in Figure \ref{fig:labSetup}. A white light source (Energetiq EQ-99XFC) was used to illuminate an integrating sphere with a 2 inch diameter exit port. Light exiting the sphere was polarized with a conventional wire grid polarizer (model PPL04C from Moxtek, Inc.), which we refer to as the ``analyzer''. The analyzer's contrast ratio was characterized, at several wavelengths, with a Wollaston prism and a calibrated photodiode. The polarizer's contrast varies from $\sim$400:1 to 2500:1, over the range of 400 - 700 nm. The measured contrast and the resulting polarization purity of the light used for characterization of micropolarizer response are given in Table \ref{tbl:polarizerContrast}. The instrument's response was characterized using broadband Bessel BVR filters (\textit{i.e.}, the passbands we used for on-sky observing).   

\begin{table}[ht]
    \centering
    \begin{tabular}{l c c} 
    \hline
         Wavelength  nm & Contrast & Polarization Purity (\%) \\
         \hline
         400 & 392$\pm$19 : 1 & 99.74$\pm$0.01 \\
         500 & 1160$\pm$25 : 1 & 99.913$\pm$0.002  \\
         600 & 1786$\pm$21 : 1 & 99.944$\pm$0.001 \\
         700 & 2473$\pm$44 : 1 & 99.959$\pm$0.001 \\
    \end{tabular}
    \caption{The contrast ratio of the analyzer (model PPL04C) used to produce polarized light for the characterization of RITPIC and the corresponding polarization purity of the light.}
    \label{tbl:polarizerContrast}
\end{table}

This setup produces polarized ``flat field'' images. A motorized rotation stage was used to change the analyzer's orientation. Multiple frames were acquired at each orientation to filter out cosmic rays (using median clipping) and to improve signal-to-noise ratio (especially for pixels orthogonal to the ``analyzer'' polarizer's orientation, that receive very little light). By acquiring polarized flats over a wide range of polarization angles, the response of every pixel (to polarized light) can be determined. 

\begin{figure}[ht]
\centering
\includegraphics[width=\linewidth]{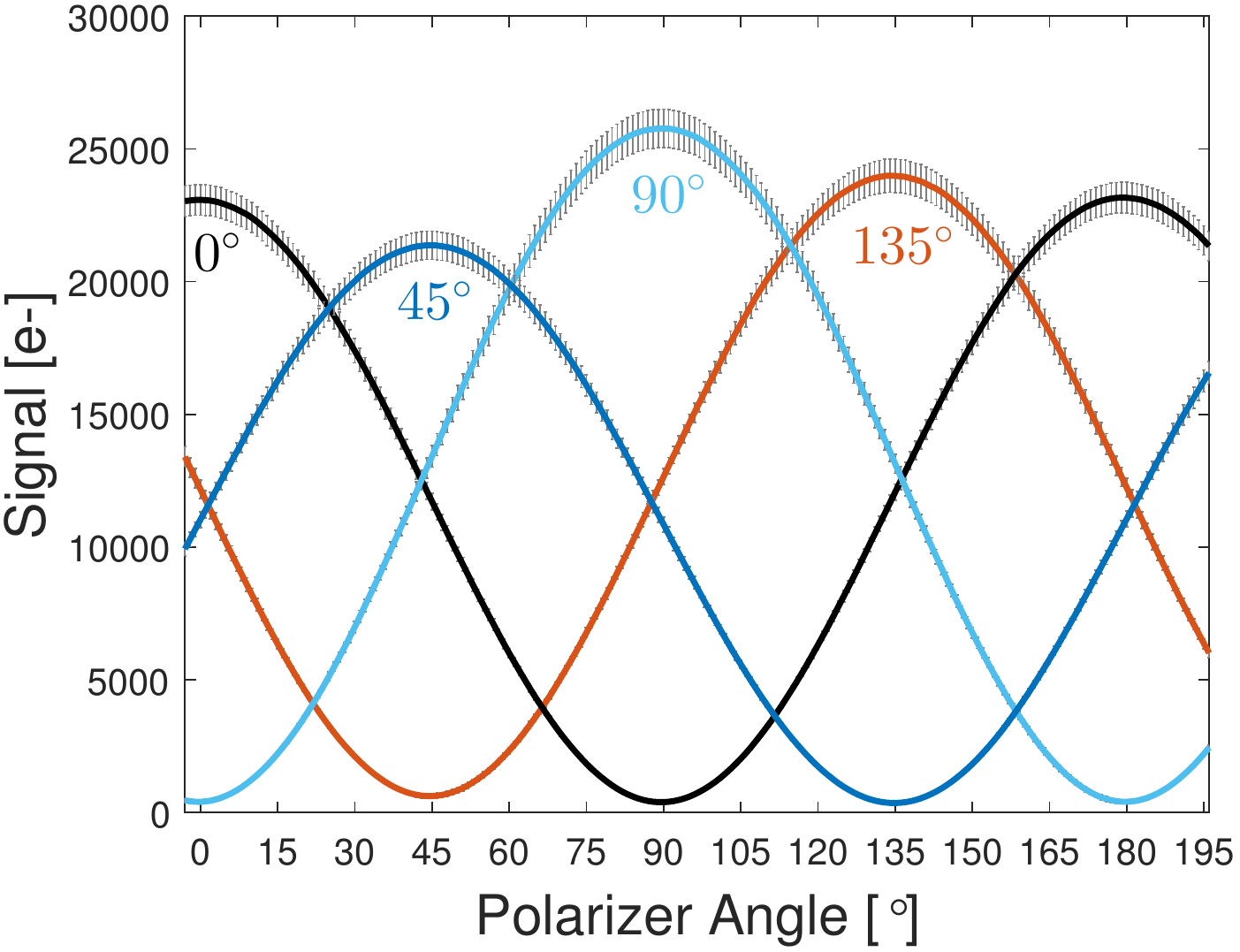}
\caption{The median response for pixels oriented along 0$^{\circ}$, 45$^{\circ}$, 90$^{\circ}$, and 135$^{\circ}$ to polarized light with a range of angles, through the Bessel R filter. The error bars show the standard deviation of pixels with the same orientation, at a particular angle.}
\label{fig:rawResponse}
\end{figure}

An example of raw response curves acquired using the Bessel R filter are shown in Figure \ref{fig:rawResponse}. It is possible to describe some properties of the system, qualitatively, from these raw curves alone. For example, the $45^{\circ}$ pixels appear to have a lower overall throughput than pixels with the other three orientations. The response curves appear offset from each other in increments of roughly 45$^{\circ}$. Also, the ratio between the minimum and maximum signal measured for pixels of a certain orientation is indicative of the contrast or efficiency of these polarizers (and it ranges from $\sim$ 30:1 - 40:1).

Once the response of each pixel to polarized light has been measured, the properties of each pixel can be determined using a model of the system. The signal measured by each pixel for light of a certain polarization can be described \citep{Sparks1999, Hines2000} with Equation \ref{eq:reductionEquation},
\begin{equation}
\label{eq:reductionEquation}
    S_k = A_k I + \epsilon_k(B_k Q + C_k U),
\end{equation}
where $S_k$ is the measured signal, $I$, $Q$, and $U$ are the Stokes parameters which describe the state of the incident light, $A_k$, $B_k$, $C_k$, and $\epsilon_k$ are transmission and efficiency terms that describe each polarizer and $k$ denotes the polarizer orientation. Using the relations between degree of polarization and the stokes parameters, the above equation can be expressed as follows,
\begin{equation*}
    S_k = A_k I + \epsilon_k(B_k I p \cos{2 \psi} + C_k I p \sin{2 \psi}),
\end{equation*}
where $I$ is the total intensity, $p$ is the intrinsic fractional polarization of the source, and $\psi$ is the intrinsic polarization angle. Next, the transmission terms $A_k$, $B_k$, and $C_k$ are written in terms of a generic throughput, $t_k$ and polarizing efficiency $\epsilon_k$,

\begin{align*}
    S_k = I \frac{1}{2} t_k & + \frac{1}{2} t_k \epsilon_k I p \cos{2\phi_k} \cos{2 \psi} \\ 
                            & + \frac{1}{2} t_k \epsilon_k I p \sin{2 \phi_k} \sin{2 \psi}.
\end{align*}

We simplify further by collecting terms and using value of $p=1$ to arrive at the system response equation,
\begin{equation}
\label{eq:polarizerModel}
    S_k = \frac{1}{2} t_k I\Big[1 + \epsilon_k (\cos{2\phi_k} \cos{2 \psi} + \sin{2 \phi_k} \sin{2 \psi})\Big].
\end{equation}
This model can be fit to the response curves in Figure \ref{fig:rawResponse} to determine the transmission, $t_k$, efficiency, $\epsilon_k$, and orientation, $\phi_k$, of every polarizer. To determine the value of $t_k$ and $\phi_k$ absolutely, the incident flux, $I$ and angles $\psi$ must be known with high accuracy. These absolute measurements are needed to determine the Stokes parameters, $I$, $Q$, and $U$. However, the measurement of the normalized Stokes parameters, $q=Q/I$ and $u=U/I$, fractional polarization, $p=\sqrt{q^2+u^2}$ and the angle of linear polarization, $\psi$, only needs the relative transmissions and orientations between the polarizers. In this case, the absolute transmission cancels out, and the $t_k$ terms act like a conventional flat field correction. Similarly, the absolute orientation of the polarizers does not matter - only the relative orientation between polarizers is important. For the sake of brevity, we only show the results acquired through the Bessel R filter. The full multi-wavelength characterization is given in Appendix \ref{sec:appendixMultiwavelengthCharacterization}.

\subsection{Pixel Throughput}
The throughput of each polarimeter pixel for unpolarized light, $t_k$, is difficult to measure accurately; this is an absolute measurement, akin to the quantum efficiency of a detector. Using a calibrated photodiode, we measured the throughput for RITPIC's micropolarizer array to be $28\pm4\%$, in the Bessel V band; we do not expect the throughput to change dramatically across the visible range. This measurement roughly agrees with a simple theoretical estimate (for an unpolarized beam),

\begin{align*}
    t = [I_p t_p + I_s(1/c)]t_{ff} &= [0.5\times 0.8 + 0.5\times (1/30)] \\ 
    \times 0.8 &= 0.333, 
\end{align*}

where $t$ is the throughput to unpolarized light, $I_p$ is the intensity in the $p$ state, $t_p$ is the transmission of the polarizer for the $p$ state light, $I_s$ is the intensity in the orthogonal $s$ state, $c$ is the contrast (the ability to block the $s$ state), and $t_{ff}$ is the throughput due to the fill factor of the micropolarizer array (estimated using reflection microscopy images of micropolarizer arrays). 

\begin{figure}[ht]
\centering
\includegraphics[width=\linewidth]{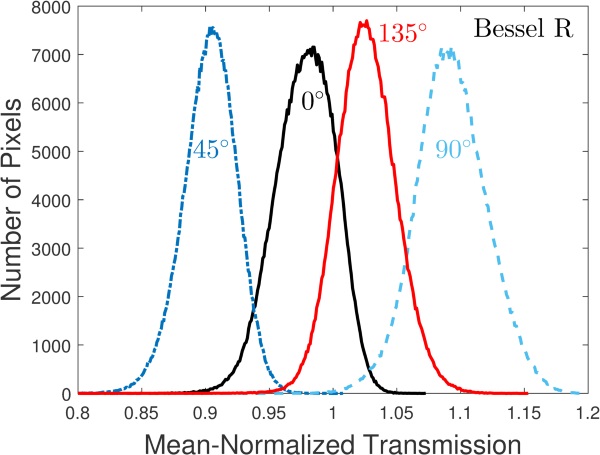}
\caption{Histograms of the mean-normalized transmissions, $t_k$, for pixels of each orientation. The throughput is systematically different for pixels with different orientations. There is a broad distribution of transmission even for pixels of the same orientation.}
\label{fig:throughput_histogram}
\end{figure}

Unlike the absolute throughput, the relative throughput differences between pixels must be known precisely, to minimize polarimetric errors. We determine the relative throughput of each pixel by mean-normalizing the $t_k$ from the best fit model (Eq \ref{eq:polarizerModel}). The mean-normalized throughput of all pixels is shown in Figure \ref{fig:throughput_histogram}). The throughput of RITPIC pixels depends systematically on their orientation. Additionally, even pixels with the same orientation show a considerable amount of variability. 

\vspace{10pt}

\begin{figure}[ht]
\centering
\includegraphics[width=\linewidth]{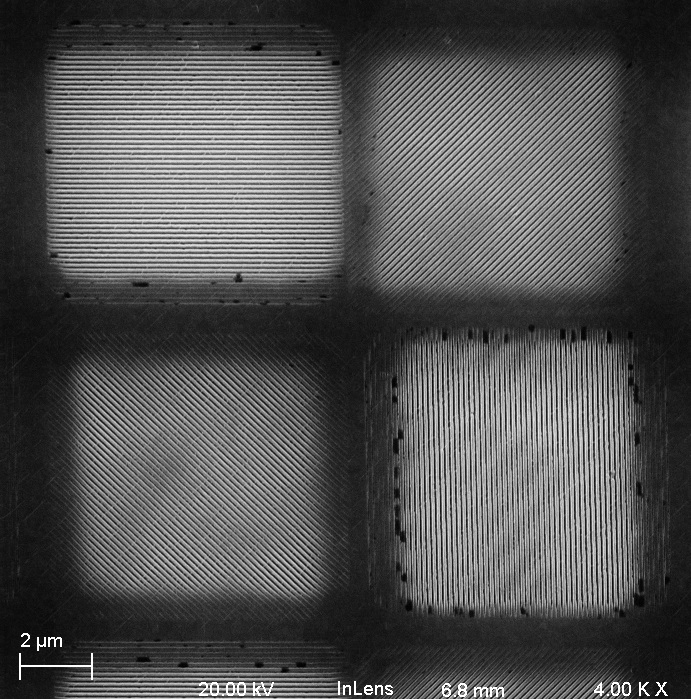}
\caption{A micrograph from a scanning electron microscope of a micropolarizer array ``superpixel''. The active area of the micropolarizer oriented along $-45^{\circ}$ is noticeably smaller than the horizontally oriented micropolarizer above it. This pattern is repeated across the array, giving rise to systematic differences in throughput between pixels of different orientations.}
\label{fig:micropolarizerSEM}
\end{figure}

\begin{figure}[ht]
\centering
\includegraphics[width=\linewidth]{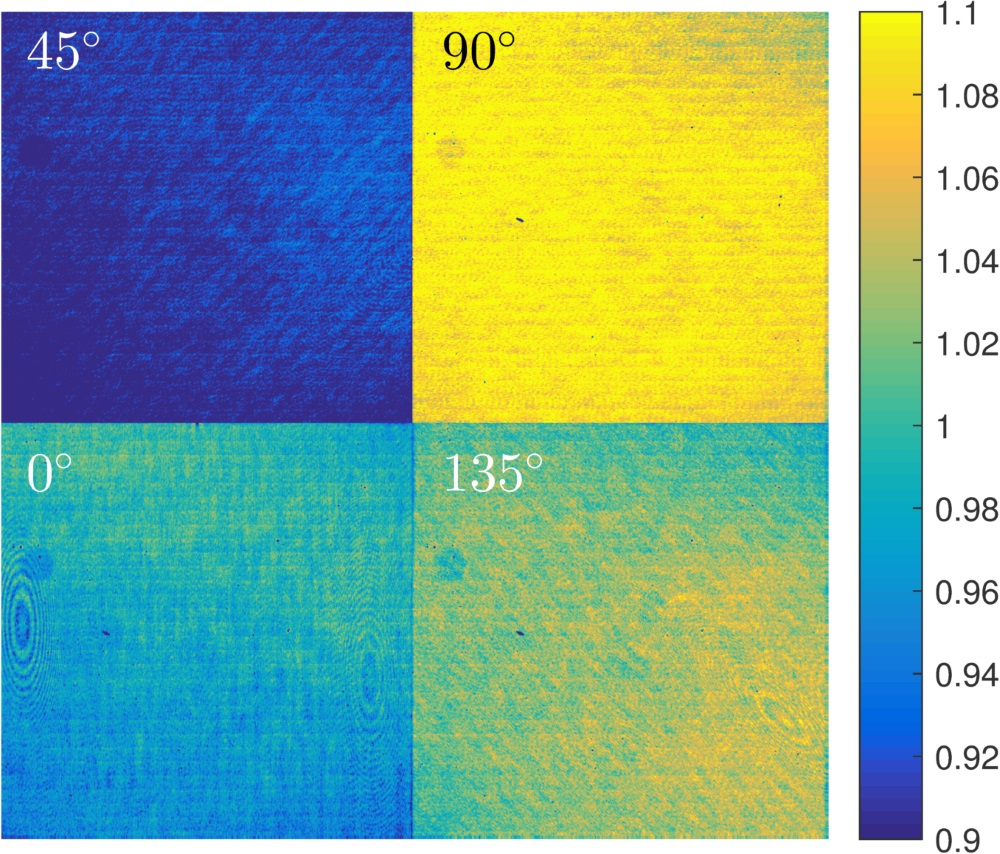}
\caption{A composite image showing the relative throughputs, $t_k$, for each pixel orientation, normalized by the mean throughput of all pixels. The systematic differences between pixels, as well as the throughput differences across the array, are clearly seen. Note, the image of the bottom-right micropolarizer is showing aliasing; the actual wires are quite straight.}
\label{fig:spatialDistribution_tk}
\end{figure}

The throughput differences between pixels arise from the fabrication process used to make the micropolarizer array and from imperfect alignment between the micropolarizer array and the imaging sensor. The systematic throughput difference between pixels of different orientations is due to systematic differences in micropolarizer effective area. This effect is a result of the fabrication process used for older generations of micropolarizer arrays (which were used to make RITPIC) and can be seen in scanning electron micrographs of individual micropolarizers (Figure \ref{fig:micropolarizerSEM}). In addition to these systematic differences, micropolarizers with the same orientation show a considerable intrinsic throughput variability. This occurs for two reasons. First, the fabrication process is not uniform across the wafer (from which each micropolarizer array is diced) and some pixels end up with fabrication defects of one type or another. Second, imperfect alignment between the micropolarizer grid and the pixel grid of the imaging sensor results in a range of ``effective throughputs'' across the resulting device, because some pixels will be better aligned than others and will couple their light more efficiently. These large-scale variations in throughput can be seen in the flat field images (Figure \ref{fig:spatialDistribution_tk}). Recent generations of micropolarizer arrays show much better response uniformity than the devices used for RITPIC. 

\subsection{Polarizer Efficiency}
The efficiency of the polarizer, $\epsilon_k$, describes its ability to reject the unwanted polarization states. The efficiency can be expressed using a ``leak'' term, $l_k$, which is the throughput of the polarizer for light oriented perpendicular to the polarizer's axis, 
\begin{equation*}
    \epsilon_k = \frac{1-l_k}{1+l_k}.
\end{equation*}

An efficiency of 0.94 corresponds to a leak of $l_k=0.03$. This roughly agrees with a contrast of $\sim$30:1. This low contrast appears to be due to cross-talk between nearby pixels, as the efficiency of an individual pixelated polarizer is significantly higher.  

\begin{figure}[ht]
\centering
\includegraphics[width=\linewidth]{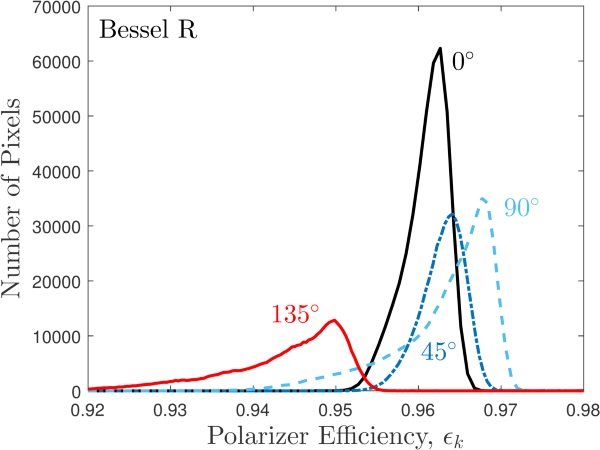}
\caption{Histograms of the polarizer efficiency, $e_k$, for pixels of each orientation.}
\label{fig:histogramsE_k}
\end{figure}

The systematic differences in efficiency of RITPIC pixels are not as dramatic as those seen in the throughput, $t_k$; except for 135$^{\circ}$ pixels. Furthermore, the range of efficiency for pixels of the same orientation are more narrow than that of $t_k$ (Figure \ref{fig:histogramsE_k}). Curiously, the pixels with highest throughput do not always correspond to pixels with highest efficiency. This may be because areas of low throughput also show very low amounts of ``leaked'' light (see Appendix \ref{sec:appendixMultiwavelengthCharacterization} for the full characterization).   

\subsection{Pixel Orientation}
\begin{figure*}[ht]
\centering
\includegraphics[width=\textwidth]{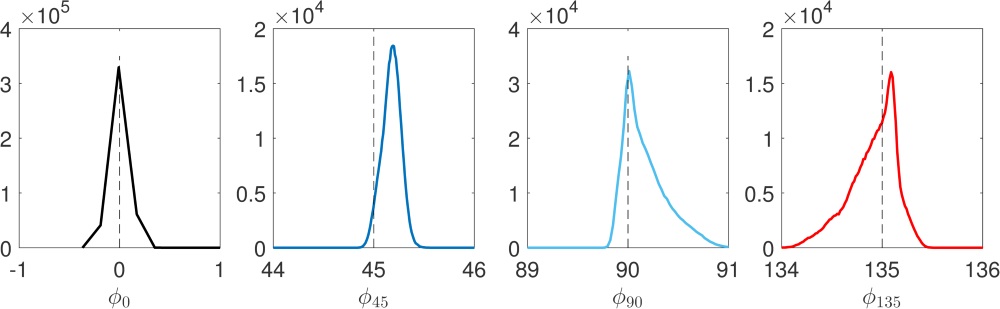}
\caption{Histograms of the polarizer orientations, $\phi_k$, for pixels of each orientation.}
\label{fig:histogramsP_k}
\end{figure*}

The last parameter needed to describe a polarizer is its orientation, $\phi_k$. In polarimetry, the relative orientations between polarizers must be known with high accuracy, because the orientation contributes to the transmission coefficients $B_k$ and $C_k$ in equation \ref{eq:reductionEquation}. Fitting the polarizer model to the response curves allows one to determine the relative orientations between pixels with high precision. The distributions of orientations for each pixel orientation are shown in Figure \ref{fig:histogramsP_k}.

On average, RITPIC's pixels are close to their nominal orientations. However, some pixels show relatively broad distributions of orientation. For example, there are hundreds of $90^{\circ}$ and $135^{\circ}$ pixels that are $1^{\circ}$ away from their nominal position. 

These orientations may not exactly reflect the physical direction of the micropolarizer wires. Instead, they are an ``effective'' orientation. That is, these polarizers behave as having these orientations. This is important to note, because the effective orientation can be affected by the alignment between the micropolarizer and the CCD (which is never perfect). Also, this means the effective orientation can change across the sensor (see Appendix \ref{sec:appendixMultiwavelengthCharacterization}). 

\subsection{Summary of Device Characterization}
We carried out characterization of RITPIC using a polarized broadband source, to determine the relative throughput, polarizer efficiency, and relative orientation of RITPIC's pixels. These metrics were obtained by fitting an instrumental model (Eq. \ref{eq:polarizerModel}) to the raw pixel response, using the Bessel BVR filters. We summarize RITPIC's performance as follows:

\begin{enumerate}[topsep=1ex,itemsep=-1ex,partopsep=1ex,parsep=1ex]
    \item Throughput for unpolarized light is $28\pm4\%$ in the BVR passbands.
    \item Pixels of different orientations show systematic throughput differences (Figure \ref{fig:throughput_histogram}).
    \item Pixels of the same orientation exhibit a throughput difference as large as $\sim10\%$.
    \item Throughput differences show large scale structure in the flat field images.
    \item Typical polarizer efficiency of RITPIC pixels is $\sim0.96$ (Figure \ref{fig:histogramsE_k}).
    \item The systematic differences in efficiency are less dramatic than in throughput.
    \item On average, RITPIC's pixels are separated by 45$^{\circ}$ $\pm$ 0.5$^{\circ}$ (Figure \ref{fig:histogramsP_k}).
\end{enumerate}

The relative throughput, efficiency, and orientation are wavelength-dependent. We stress that the precise estimation of these parameters is critical for precise polarimetry. A full description of the multi-wavelength characterization of RITPIC can be found in Appendix \ref{sec:appendixMultiwavelengthCharacterization}.  

\section{Data Acquisition and Observing Strategy} \label{sec:dataAcquisition}
Polarimeters based on micropolarizer arrays (wire grid or polymer) represent a new subset of the division of focal plane modulation strategy. The most distinguishing feature of MPA-based polarimeters is the sampling of different parts of the image by polarizers of different orientations. This means that each pixel's ``instantaneous field of view'' (\cite{Tyo2009}) is unique. In turn, the Stokes parameters cannot be estimated directly from the intensities measured by a set of four polarizers with orientations 0$^{\circ}$, 45$^{\circ}$, 90$^{\circ}$ and 135$^{\circ}$, as described in equation \ref{eq:reductionEquation}, because each pixel might be sampling a portion of the image that has an intrinsically higher intensity or different polarization properties. Furthermore, cross-talk present in the detector can introduce systematic errors during the data analysis stage. In this section we discuss issues related to proper sampling.

\subsection{Proper Sampling and the Instantaneous FOV} 
\label{sec:samplingAndIFOV}
Division-of-focal-plane polarimeters modulate the polarization state across the scene (which is imaged onto the focal plane) on a pixel-by-pixel basis. A set of 4 pixels with different orientations, a ``superpixel'', is the modulation element of the polarization-sensitive array. The polarization of the incident light is reconstructed by de-modulating the intensities recorded by each pixel. Literally, it means that a division-of-focal plane polarimeter should be designed such that the scene imaged onto the polarization-sensitive array not have spatial frequencies higher than the sampling frequency of the superpixels of the array; however, it is unlikely that this can ever be practically implemented. As such, appropriate data-reduction techniques must be developed to minimize the effects of the instantaneous field of view differences.

This problem was identified by several groups, who used various techniques to combat the false polarization signals that it tends to cause (\cite{Tyo2009}, \cite{Gao2011}). Although these approaches seem to reduce visual artifacts during the polarization analysis, it is difficult to estimate the effect on the accuracy and precision of the polarimetric estimation. An in-depth discussion of the effects of sampling can be found in \cite{Vorobiev2017phd}. Here, we summarize some key aspects associated with observations of point sources and extended objects.

\subsubsection{Observations of Unresolved Objects}
Polarimeters based on micropolarizer arrays have inherent imaging capability. However, they can also be used for polarimetry of unresolved sources, using conventional photometric techniques (such as aperture photometry, with or without PSF-fitting). Images of stars and other point sources show the most dramatic differences in instantaneous field of view. By definition, the PSF has the most steep intensity gradients possible in an image. Analysis of synthetic observations presented in \cite{Vorobiev2017phd} shows that the PSF should be sampled by 4 - 5 ``superpixels,'' to minimize polarimetric errors caused by non-uniform illumination of the micropolarizer array. When the PSF is sampled by fewer pixels than this, the individual pixels within a superpixel are illuminated by steep intensity gradients, which undermines the arithmetic of polarimetry (see Section \ref{sec:velax1}). 

\begin{figure*}[ht]
\centering
\includegraphics[width=0.8\textwidth]{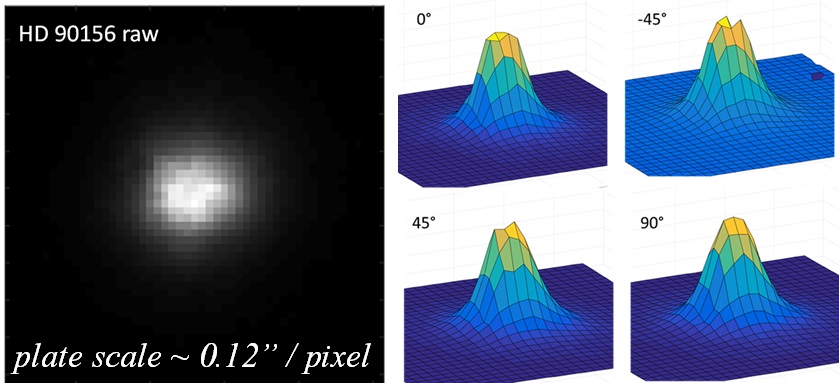}
\caption{Left: a raw image of the unpolarized standard star HD 90156. Right: once the full image is divided into subframes with pixels of the same orientation, the flux can be measured using conventional aperture photometry. The seeing was $\sim$1\arcsec.}
\label{fig:hd90156}
\end{figure*}

\subsubsection{Observations of Extended Objects}
In theory, observations of extended objects (``imaging polarimetry'') is subject to the same sampling constraints as the case of point sources. In practice, \textit{a priori} knowledge of the target can be used to slightly relax the sampling constraints. For example, various nebulae often lack sharp features, which significantly reduces the intensity of gradients at the focal plane. Similarly, the disks of solar system planets rarely show point-like features (other than the planet's limb). If errors associated with edges can be tolerated, then the sampling could be reduced (for example, to increase field of view).

\section{Data Analysis and Calibration}

Raw data from RITPIC can be processed using aperture photometry techniques (for unresolved sources) or a pixel-by-pixel approach, which is required for extended objects. In this section we describe these two methods, with a special emphasis on the pixel-by-pixel ``imaging'' approach. 

The data reduction process for RITPIC closely follows traditional CCD techniques. In both cases, the raw counts from the CCD must be corrected for the bias offset, dark current, and CCD gain. The calibration procedure begins to deviate from the typical case when the pixel response non-uniformity needs to be accounted for. This is handled differently for the case of point sources and extended objects.

\subsection{Polarimetry of Unresolved Sources}
\label{sec:polarimetryUnresolvedSources}
Once the raw counts have been dark-subtracted and converted to electrons, the raw intensities measured by each pixel have to be corrected for the effects of pixel response non-uniformity. As we mentioned in Section \ref{sec:performanceCharacterization}, the response of each pixel depends on its throughput to unpolarized light, its polarization efficiency, and its orientation. Unlike conventional CCDs, pixels of different orientations show systematic differences, in addition to some intrinsic variability. We correct each of these effects in separate steps.

First, the calibrated signal frames (in units of electrons) are separated into subframes, with each frame composed of pixels with the same orientation. An example raw frame and its corresponding subframes for the unpolarized standard HD 90156 are shown in Figure \ref{fig:hd90156}. Aperture photometry can then be performed in each subframe to estimate the total target flux in each subframe.  

Once the target flux in each subframe has been measured, these values can be demodulated to determine the Stokes parameters that describe the polarization of the light. To account for the imperfect polarizer efficiency and orientation, the demodulation procedure must incorporate the micropolarizer properties determined during a full characterization of the device. 

In this case, the ``flat field'' correction is performed as part of the demodulation process, using equation \ref{eq:reductionEquation}. The systematic differences between pixels of different orientations can be corrected using the average values of $t_k$, $e_k$, and $\phi_k$ as follows (where the subscript $k$ indicates the polarizer orientation),
\begin{equation}
\label{eq:reductionEquationMeans}
    S_k = \frac{1}{2} \langle t_k \rangle I\Big[1 + \langle \epsilon_k \rangle (\cos{2\langle \phi_k \rangle} \cos{2 \psi} + \sin{2 \langle \phi_k \rangle} \sin{2 \psi})\Big],
\end{equation}
where the angle brackets $\langle$ $\rangle$ denote an average. An average value is used here, because the Stokes parameters are determined by demodulating the total flux measured in a subframe, using data from many pixels. However, this process does not take into the differences between pixels of the same orientation, and can lead to systematic errors when stars fall on pixels whose properties are sufficiently different than the mean. The pixel-to-pixel variations of throughput in the same subframe can be calibrated by normalizing the pixel values by a flat field made using the mean-normalized maps of throughput. This normalization allows one to use the mean value of the throughput in equation \ref{eq:reductionEquationMeans}, without becoming vulnerable to the pixel-to-pixel variations. Once the raw fluxes, $S_k$, have been calibrated, the system of equations given by Eq. \ref{eq:reductionEquationMeans} can be solved using linear least squares regression. An in-depth description of point-source polarimetry with micropolarizer arrays and the associated systematic effects can be found in \cite{Vorobiev2017phd}.
\vspace{1cm}
\subsection{Polarimetry of Extended Objects}
\label{sec:polarimetryOfExtendedObjects}
In this section, we describe a process that can be used to perform true imaging polarimetry, using a per-pixel calibration, to determine the polarimetric properties for every pixel in an image. As for point sources, the first step in the demodulation process is separating the raw image into four subframes, based on pixel orientation. Next, each subframe must be flat-fielded to account for the pixel-to-pixel variations in throughput among pixels of the same orientation. This is done using mean-normalized throughput values, $t_k$, as determined during device characterization, or using a good flat field image acquired using unpolarized light. 

\subsubsection{Spatial Registration of Subframe Images}
Because each pixel is physically offset from its neighbors in the superpixel, the object in the four subframes is also shifted by 1 pixel (or 0.5 superpixels). This is another way the instantaneous field-of-view problem manifests itself. To account for this, we choose one of the subframes to act as a reference frame and spatially register the remaining subframes. Depending on the exact micropolarizer configuration, the subframes must be moved $\pm$0.5 pixels in the \^{x} and/or \^{y} directions. To perform this sub-pixel shift, we first interpolate the images into a grid with finer sampling, shift an integer number of these smaller pixels, and downsample the image to the original size.

Once the images are registered, we demodulate the intensities using the linear least squares method, on a per-pixel basis (following the approach described in \cite{Sparks1999} and \cite{Hines2000}). Practically this is done as a loop over all pixels in the subframe image. A description of the calibration process and the sorts of errors that arise during imaging polarimetry is given in Appendix \ref{sec:appendixImagingPolarimetry}.

\subsection{Sources of Uncertainty in Micropolarizer-based Polarimetry}
\label{sec:noiseSources}
Having described the data analysis process, we turn our attention to mechanisms that affect the polarimetric sensitivity and polarimetric accuracy of measurements made with micropolarizer-based polarimeters. In this work, polarimetric ``sensitivity'' refers to the amount of scatter in the estimations of the Stokes parameters. This noise floor is indicative of the smallest fractional polarization that can be detected. On the other hand, polarimetric ``accuracy'' is more affected by systematic errors that stem from instrumental effects, such as induced polarization, depolarization and cross-talk. 

In general, high quality polarimetry requires high quality photometry; therefore, many aspects of photometric analysis are relevant to polarimetry. However, because polarimetry is performed by comparing several intensities, there exist idiosyncrasies not present in conventional photometry; mechanisms that affect a sensor's dynamic range are particularly germane. Here, we briefly review some aspects of CCD photometry and explain their effects on polarimetry with MPA-based sensors (from least to most important).

\subsubsection{Read Noise} When the number of photo-electrons in a single pixel is sufficiently small, read noise can make up a significant fraction of the total uncertainty. Typically, polarimetry is done in the shot-noise dominated regime and read noise isn't the limiting factor. However, observations of very polarized objects could result in pixels with very little flux, if those pixels are oriented so as to block most of the light from the object. In these pixels, read noise could again be important, even though the object is otherwise bright.  

\subsubsection{Dark Noise} The effect of dark current and its noise are similar to those of read noise. Although most pixels should have a similar amount of dark noise (excluding hot pixels), its effects may be more important for those pixels in a superpixel that receive the least flux (due to the polarization properties of the source). 

\subsubsection{Bad Pixels} Hot pixels and dead pixels are especially problematic for micropolarizer-based polarimetry, because even a single bad pixel within a superpixel will introduce significant error into the polarimetry. This is especially true for imaging polarimetry and less severe in the case of aperture photometry. In either case, bad pixels should be identified and handled in some way. The most conservative approach is to not use the affected superpixels at all; however, interpolating over the bad pixels may also be appropriate, depending on the scene. 

\subsubsection{Charge Diffusion} Some imaging sensors can exhibit significant charge diffusion. Carriers can diffuse into the potential well of a neighboring pixel or out of the potential well completely and recombine. The latter is analogous to a decrease in the detector QE and, as a result, the photometric SNR. In conventional photometry, the former leads to a broadening of the PSF; however, as this process is carrier-conservative, it is largely a nuisance. For micropolarizer-based polarimeters, the diffusion of carriers from one pixel to its neighbor is catastrophic, because polarimetric properties of the source are determined from the difference in flux measured by neighboring pixels. In general, this kind of cross-talk tends to wash out intensity differences between pixels, leading one to underestimate the fractional polarization of the source. This effect becomes more significant as the fractional polarization increases. The amount of diffusion in a RITPIC pixel has not yet been measured directly.

\subsubsection{Charge Transfer Efficiency} A ``smearing'' of charge between neighboring pixels, similar to diffusion, can result from charge transfer inefficiency. This process and its effects on photometry are well known. In practice, CTE does not contribute significantly to photometric uncertainty, because modern CCDs have CTEs $>$0.999999 and because this process is usually carrier-conservative. However, if the CTE is sufficiently low (due to sensor design or radiation damage), it could become important in micropolarizer-based sensors.  

\subsubsection{Sensor Linearity} Modern CCDs exhibit linear response ($<1\%$ deviation from linearity) over a large fraction of the full well. Errors due to nonlinearity typically arise when comparing two sources, of which one is close to saturation. This situation can also occur in the case of micropolarizer-based polarimeters. Similarly, special attention to linearity must be paid when creating ``high dynamic range'' composite images, from a range of exposure times. Again, micropolarizer-based images may also contain superpixels with very different signal levels among its constituent pixels, for very polarized sources. The brighter pixels in these images may be close to saturation, even though their immediate neighbors are well within the linear regime. RITPIC's sensor has particularly good linearity across the entire full well, with a maximum deviation from linearity of 0.04\%.    

\subsubsection{Pixel-to-pixel Response Nonuniformity} Because RITPIC is a single-beam instrument without a modulating retarder, the intensity in orthogonal polarization channels is measured by different pixels, with slightly different responses; this is in contrast to the dual-beam dual-difference method, where the same detector can measure both channels. CCD pixels do not exhibit uniform sensitivity across the array, due to a combination of intrinsic sensor properties and outside factors, like vignetting and dust. The intrinsic sensor nonuniformity is usually on the order of 1\% and is corrected by flat fielding techniques; a detailed discussion of this fixed-pattern noise can be found in Section 4.3 of \cite{Janesick2001}. The nonuniformity of micropolarizer arrays (Section \ref{sec:performanceCharacterization}) adds to the existing sources of fixed-pattern noise. For RITPIC, these combined effects produce a nonuniformity is $\sim$10\% for pixels of the same orientation, with even bigger systematic differences between pixels of different orientations. When demodulating micropolarizer-based data on a per-superpixel basis (imaging polarimetry), the response differences of the four pixels in a superpixel must be characterized and used in the demodulation process. For RITPIC, this is performed during the lab-based characterization process, by determining the throughput, efficiency, and orientation of every pixel (Section \ref{sec:performanceCharacterization}). When performing aperture photometry, a technique analogous to conventional flat fielding must be employed, because the response of many pixels is used to determine the total flux in that subframe. 

\subsubsection{Instrumental Polarization} Every optical element has the potential to increase or decrease the fractional polarization, to rotate the angle of linear polarization, and/or create cross-talk between the linear and circular components; in general, the term ``instrumental polarization'' refers to all of these effects. In practice, one strives to keep the instrumental signature minimal and stable. In this case, the instrumental effects can be measured with observations of polarimetric standards and calibrated (see \cite{Wiktorowicz2014}; \cite{Harrington2017} and references within). If the instrumental signature changes with time or instrument orientation, the calibration process becomes more difficult and the instrumental polarization may decrease the polarimetric sensitivity. Instrumental effects are especially important to micropolarizer-based sensors operating in a true ``snapshot'' mode, without an additional modulator, because they cannot take advantage of the dual-difference technique, which helps mitigate some of these effects.    

\subsubsection{Instantaneous Field-of-View} Division-of-focal plane sensors sample the image in the spatial and polarimetric domains simultaneously, as described in Section \ref{sec:samplingAndIFOV}. As such, aliasing can arise from high spatial frequencies in the intensity or polarimetric content. If the image sampled by the micropolarizer sensor is sufficiently bandlimited, the raw intensities can be demodulated without introducing additional uncertainty into the polarimetry \citep{Tyo2009}. However, truly bandlimited observations are not always practically possible or desirable. In these cases, careful attention must be paid to the sampling at the focal plane. The optimal sampling rate will depend on the modulation transfer function of the optics, any atmospheric effects, and the spatial frequency content of the scene. As a general guideline, the PSF should be sampled with $\sim$3 - 4 superpixels. However, the overall uncertainty also depends on the photometric SNR and effects related to seeing (if any); a more in-depth discussion of these effects can be found in \cite{Vorobiev2017phd}. 

\subsubsection{Summary - Sources of Uncertainty} 
In this subsection, we attempted to identify mechanisms which might contribute to the uncertainty of polarimetric measurements made with micropolarizer-based sensors. Which of these will be the dominant source of uncertainty in practice depends on the properties of each polarimeter, on the observing conditions, and the properties of the source. In most cases, polarimetry is done with many photons, and effects of read noise and dark noise are small. When the flux measured in each sub-pixel is shot-noise dominated, the overall polarimetric sensitivity and accuracy are limited by the quality of the calibration of systematic effects and uncertainty related to proper sampling (IFOV errors). In the next section, we estimate RITPIC's sensitivity using observations of polarimetric standards. 

\section{RITPIC's On-Sky Performance}
The current generation of the RIT Polarization Imaging Camera (RITPIC) was deployed on a Boller \& Chivens 36'' telescope at the Cerro Tololo Inter-American Observatory (CTIO) 2016  February 3 - 13. during our run, 7 nights could be considered ``photometric'', with seeing varying from 0.75'' - 1.5''; in this work, we only present the results obtained on these nights. We observed a wide range of objects: calibration stars, planetary nebulae, post-asymptotic giant branch stars, Solar System planets, protoplanetary nebulae, open clusters, globular clusters, quasars and the highly obscured supernova SN2016adj in Centaurus A. These objects were chosen to estimate the suitability of RITPIC for polarimetry of point sources and extended objects. Most of these objects have been observed by other polarimeters; however, some objects do not have polarimetric information and hold the potential for new discoveries.

The camera was mounted on the existing filter wheel, at the Cassegrain focus. Flat field illumination images were acquired using a screen in the dome and 3 lamps placed symmetrically around the telescope aperture. However, these flats were not used in the analysis presented in this work. Instead, the pixel-to-pixel nonuniformity correction was performed using the characterization performed in the lab (see Section \ref{sec:performanceCharacterization}). Therefore, instrumental effects due to the telescope optics may not be accounted in the performance characterization. As such, the results presented here may contain un-calibrated instrumental effects. In the future, we plan to develop a more robust characterization process, which will measure the performance of the entire system: telescope and polarimeter.

\subsection{Observations of Unresolved Sources}
We observed several stars whose polarimetric properties are known from previous work. Here we present our polarimetry of the weakly polarized standard HD 90156 and the polarized star Vela X-1.  

\subsection{HD 90156}
The star HD 90156 was measured by \cite{GilHutton2003} to be very weakly polarized, with $q = u = 0.00006 \pm 0.0001$. On the night of 2016 February 3, we obtained several exposures using Bessel BVR filters, using the parameters given in Table \ref{tbl:HD90156BVR}. The polarimetry was performed in ``aperture photometry mode'', following the procedure described in Sec. \ref{sec:polarimetryUnresolvedSources}.   

\begin{table*}[ht]
\begin{tabularx}{\textwidth}{ *{6}{S} } 
    \hline
    \hline
         Filter & Exposure (s) & No. of Exposures & Stokes $q$ & Stokes $u$ & DOLP\\
         \hline
         B & 10 & 10 & 0.002 $\pm$ 0.003 & -0.010 $\pm$ 0.004 & 0.010 $\pm$ 0.004 \\
         V & 5 & 10  & 0.000 $\pm$ 0.004 & 0.002 $\pm$ 0.003  & 0.002 $\pm$ 0.003 \\
         R & 3 & 20  & -0.002 $\pm$ 0.006 & -0.002 $\pm$ 0.007  & 0.003 $\pm$ 0.007 \\
         none$^{a}$ & 0.5 & 200 & 0.004 $\pm$ 0.012 & 0.003 $\pm$ 0.012 & 0.005 $\pm$ 0.012 \\
    \end{tabularx}
    \caption{Observations of HD 90156 were performed using Bessel BVR filters on the night of 2016, February 3. Note: $^a$Additional observations, without any filters, were obtained on 2016, February 8.}
    \label{tbl:HD90156BVR}
\end{table*}

\begin{figure*}[ht]
\centering
\includegraphics[width=\textwidth]{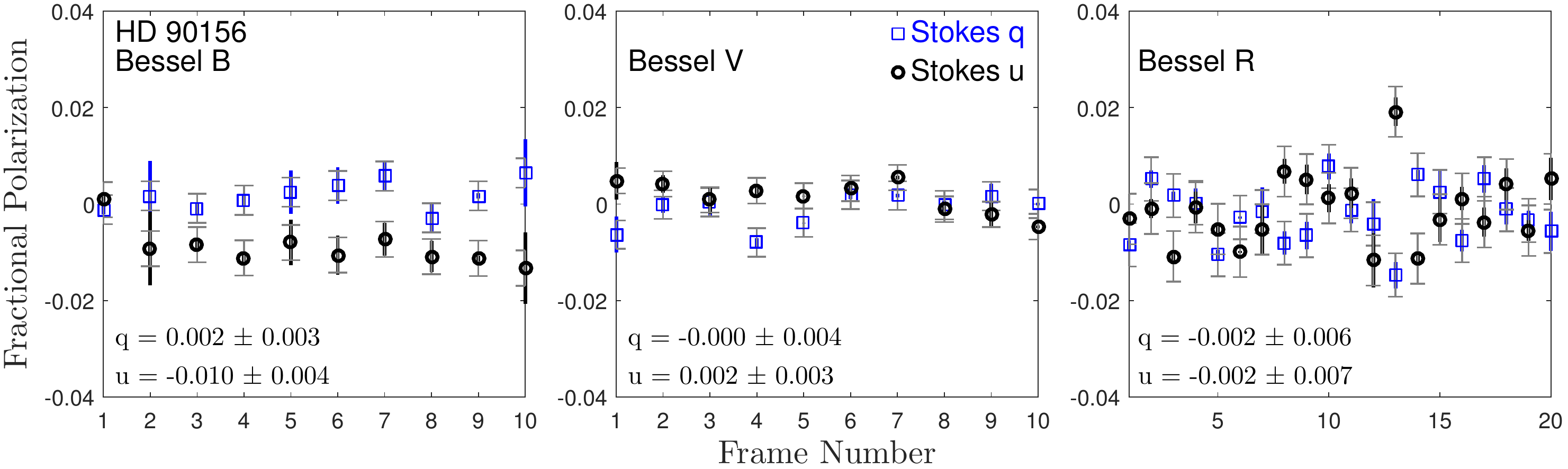}
\caption{Normalized Stokes parameters $q$ and $u$, measured by RITPIC for the unpolarized star HD 90156 in the BVR bands. The error bars with horizontal marks are estimated using the frame-to-frame scatter of the flux estimation in each subframe ($0^{\circ}$, $45^{\circ}$, $90^{\circ}$, etc.), whereas the thick vertical bars show the residual from the model fit in each measurement. The median values of $q$ and $u$ and the standard deviation are given in the lower left corner.} 
\label{fig:HD90156_BVR_polarimetry}
\end{figure*}

The Stokes parameters obtained in the BVR bands are shown in Figure \ref{fig:HD90156_BVR_polarimetry}. The results appear to show some systematic effects in each passband. In the B band, Stokes $u$ appear to show evidence of uncorrected instrumental polarization; Stokes $q$, on the other hand, is consistent with zero, within one standard deviation. The V band data appear the most well behaved, with the least frame-to-frame scatter and instrumental effects. Finally, the R band data show considerable scatter, but no obvious residual instrumental effects. A more detailed analysis of the uncetainty is presented in Appendix \ref{sec:appendixAnalysisOfUncertainty}. 

The normalized Stokes parameters, $q$ and $u$, can be used to determine the overall degree of linear polarization (DOLP) and the angle of linear polarization (AOLP), using the usual definitions \citep{hecht2002optics},

\begin{equation}
\label{eq:dolp}
DOLP = \sqrt{q^2 + u^2}
\end{equation}
and
\begin{equation}
\label{eq:aolp}
AOLP = \frac{1}{2}\textrm{atan}(\frac{u}{q}).
\end{equation}

These two metrics are commonly used to describe the polarization of an object or scene, even though they are biased estimators (see \cite{Wardle1974, Simmons1983, Simmons1985}). Despite its bias, the DOLP is useful because it is independent of any reference frame. This makes comparison of observations made by different instruments and authors more straightforward (at the cost of somewhat reduced information content). The DOLP measured for HD 90156 in the BVR bands given in Table \ref{tbl:HD90156BVR}. To help minimize the impact of noise, we use the median values $\bar{q}$ and $\bar{u}$ to calculate the DOLP and AOLP; the associated uncertainty is determined using the standard deviation of $q$ and $u$, and formal error propagation. 

\subsubsection{Vela X-1}
\label{sec:velax1}
\begin{figure*}[ht]
\centering
\includegraphics[width=\textwidth]{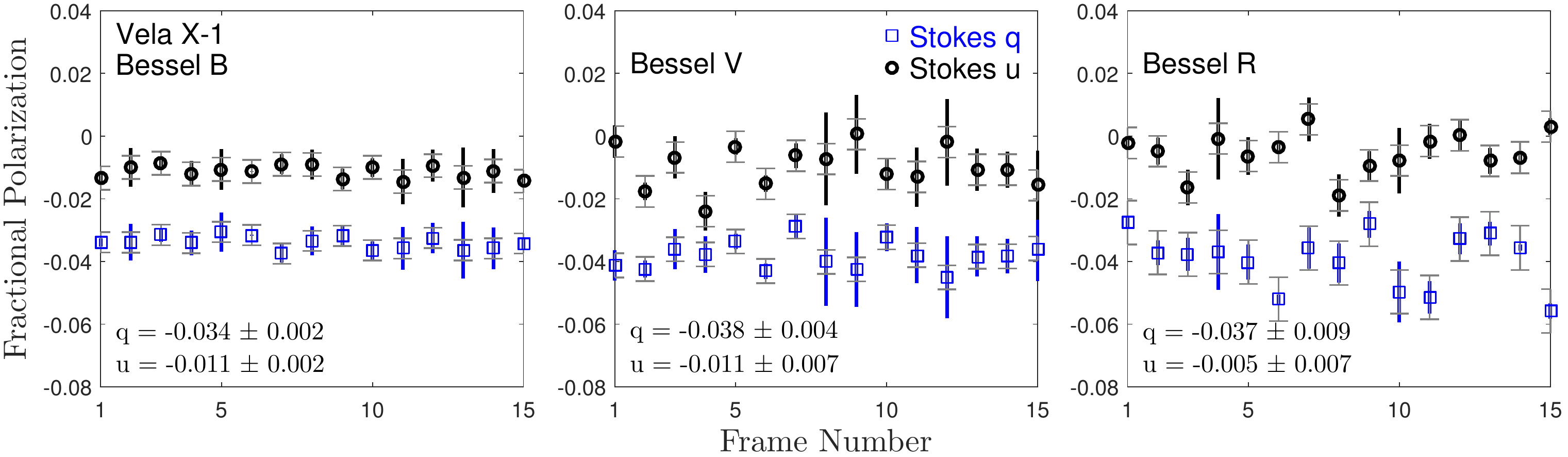}
\caption{Normalized Stokes parameters $q$ and $u$, measured by RITPIC for the polarized binary Vela X-1 in the BVR bands. The median values of $q$ and $u$ and the standard deviation are given in the lower left corner. The effects of seeing and PSF size can be seen in the difference in scatter between the B and VR observations; better seeing in the VR observations resulted in a slightly undersampled PSF.} 
\label{fig:VelaX1BVR}
\end{figure*}

Next, we present our observations of the high mass x-ray binary Vela X-1, acquired on 2016, February 4 (Figure \ref{fig:VelaX1BVR}. Details of the observations and results of the polarimetry are summarized in Table \ref{tbl:VelaX1BVR}. The fractional polarization of Vela X-1 varies with the orbital phase between $p\approx0.035 - 0.037$ in the B band, and $p\approx0.037 - 0.039$ in the V band \citep{Paradijs1980}. As such, it is difficult to use a single night of observations of Vela X-1 to estimate any instrumental effects with high precision. However, the variability of Vela X-1 is comparable to the scatter in our measurements and our polarimetry of this system is consistent with the measurements of \cite{Paradijs1980}, given our uncertainty. Also, these observations show how sensitive the frame-to-frame scatter is to sampling at the focal plane. In the B band, 97\% of the flux in each subframe is in an aperture with radius of 14$\pm$1.2 pixels. For the V and R observations, this radius is 11$\pm$1.2 pixels. A summary of measurements of these unpolarized and polarized standards is given in Table \ref{tbl:calibrationStandards}.

\begin{table*}[ht]
\begin{tabularx}{\textwidth}{ *{6}{S} }  
    \hline
    \hline
         Filter & Exposure (s) & No. of Exposures & Stokes $q$ & Stokes $u$ & DOLP\\
         \hline
         B & 8 & 15  & -0.034 $\pm$ 0.002 & -0.011 $\pm$ 0.002 & 0.036 $\pm$ 0.002 \\
         V & 3 & 15  & -0.038 $\pm$ 0.004 & -0.011 $\pm$ 0.007 & 0.040 $\pm$ 0.004 \\
         R & 4 & 15  & -0.037 $\pm$ 0.009 & -0.005 $\pm$ 0.007  & 0.037 $\pm$ 0.009
    \end{tabularx}
    \caption{Observations of Vela X-1 were performed using Bessel BVR filters on the night of 2016, February 4.}
    \label{tbl:VelaX1BVR}
\end{table*}

\begin{table}[ht]
\begin{tabularx}{\columnwidth}{X S S} 
    \hline
    \hline
      \footnotesize{Target (Filter)} & \footnotesize{DOLP}$_0$ & \footnotesize{DOLP\textsubscript{RITPIC}} \textsuperscript{$\dagger$} \\
         \hline
         HD 90156 (V) & 0.0001\textsuperscript{a} & 0.002 $\pm$ 0.003 \\
         Vela X-1 (B) & 0.035 - 0.037\textsuperscript{b} & 0.036 $\pm$ 0.002 \\
         Vela X-1 (V) & 0.037 - 0.039\textsuperscript{b} & 0.040 $\pm$ 0.004
    \end{tabularx}
    \caption{Comparison of the polarimetric analysis of two standard stars performed by RITPIC to previous measurements (DOLP$_0$), made by \textsuperscript{a}\cite{GilHutton2003} and \textsuperscript{b}\cite{Paradijs1980}. Note: \textsuperscript{$\dagger$} Median $\pm$ standard deviation.}
    \label{tbl:calibrationStandards}
\end{table}

\subsection{Polarimetry of Extended Objects}
To evaluate the performance of RITPIC in true ``imaging'' mode, we show our observations of Saturn and the protoplanetary nebulae Hen 401. Our results are given in the instrumental reference frame, which is roughly aligned with the celestial coordinate system. More specifically, $+Q$ corresponds to an electric field oscillating in the east-west direction, and $-Q$ suggests polarization along the north-south direction. Similarly $+U$ and $-U$ correspond to polarization along the $+45^{\circ}$ and $\-45^{\circ}$ directions. Finally, going from $+Q$ to $+U$ rotates the electric field vector in the counter-clockwise direction. 

\subsubsection{Saturn}
\begin{figure*}[ht]
\centering
\includegraphics[width=\textwidth]{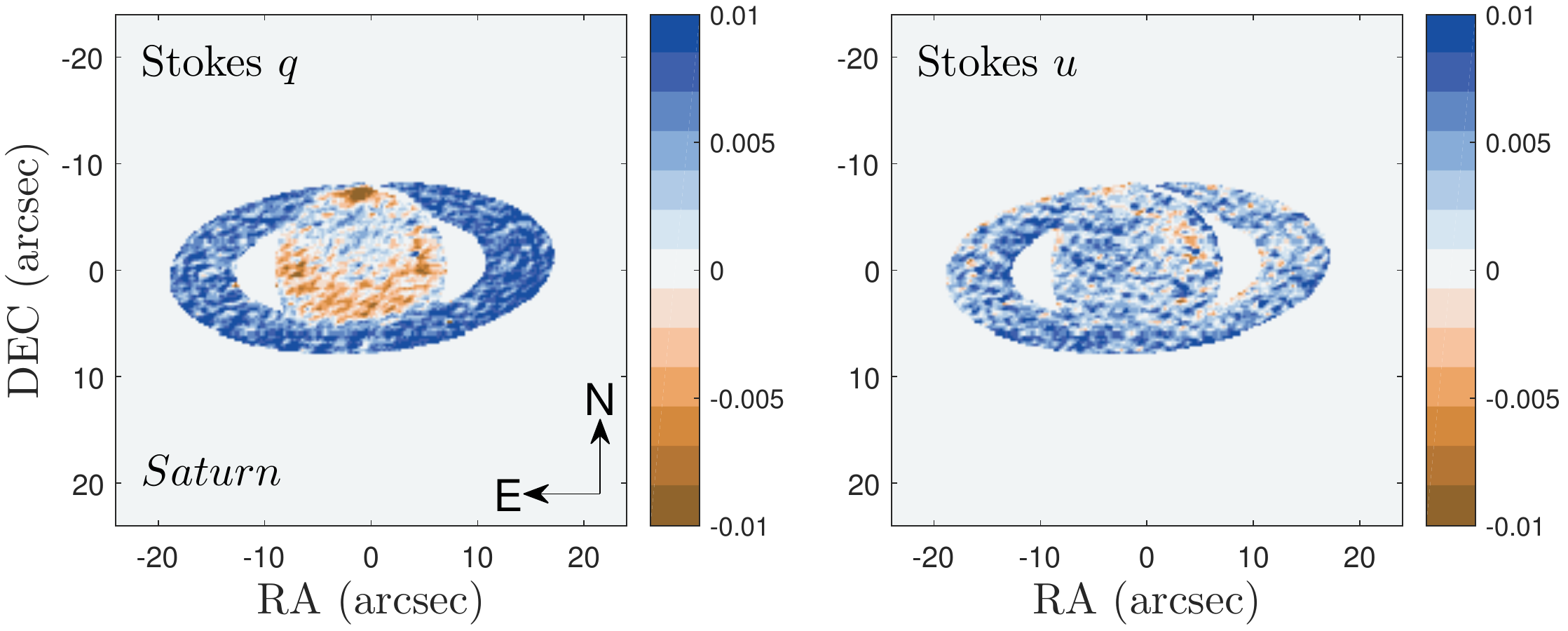}
\caption{Polarization of Saturn and its ring system in Stokes $q$ and $u$. Although the overall polarization is low, some large scale features are clearly seen, such as the polarization of the rings, the northern and southern hemisphere and the south pole.} 
\label{fig:SaturnStokesQU}
\end{figure*}

As part of the commissioning run of RITPIC, we observed Saturn on the night of February 8, 2016. Images of Saturn were acquired in the Bessel BVR bands, using 1 and 2 second exposures. The phase angle, $\alpha$ was $\sim$ 5.1$^{\circ}$. The images of nebulae were processed using the procedure described in Section \ref{sec:polarimetryOfExtendedObjects}. First, the individual subframes were registered to account for the 1 pixel offset. Next, the subframes were demodulated individually to determine the Stokes paremeters in each frame. Finally, the maps of Stokes parameters were spatially registered, to account for object motion from frame-to-frame, and combined using a median.  The maps of the normalized Stokes parameters, $q$ and $u$ are shown in Figure \ref{fig:SaturnStokesQU}.  

Saturn is weakly polarized, however, some features are clearly visible. The north pole of Saturn and the equatorial regions show negative polarization in Stokes $q$, at a level of $\sim q = -0.015$ and $\sim q = -0.005$, respectively. The higher latitude regions show a positive polarization of $\sim q=0.005$. The rings appear to have a more-or-less uniform polarization in Stokes $q$ of $\sim q = 0.01$. Both the disk and rings of Saturn show very low levels of polarization in Stokes $u$. The angle of linear polarization of Saturn shows strong, large scale patterns. The rings appear polarized at roughly $20^{\circ}$ with respect to the East-West direction. There appears to be a faint feature in the AOLP map that corresponds to the bright band near Saturn's equator. Both this band and the north pole are polarized with an angle of $\sim70^{\circ}$. This likely points to the rotation between the scattering plane and our instrumental frame.

\begin{figure*}[ht]
\centering
\includegraphics[width=0.99\textwidth]{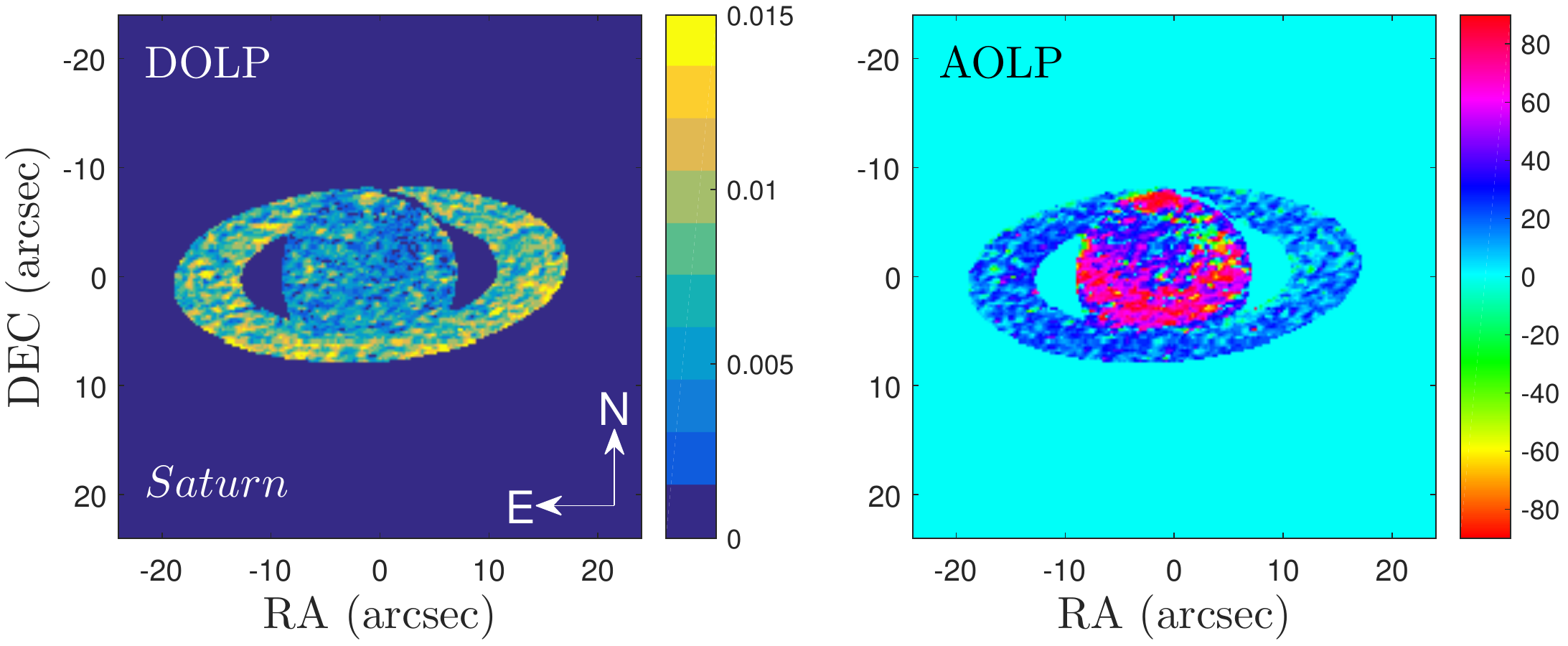}
\caption{Left: the degree of linear polarization of the disk and rings of Saturn. Right: the angle of linear polarization, in the instrumental reference frame.} 
\label{fig:SaturnDolp}
\end{figure*}

A comparison of our observations to previous work is non-trivial, because Saturn's polarization depends on the phase angle, inclination, and wavelength. Nevertheless, we see many of the features observed by \cite{Schmid2011} and \cite{Keller2002}. In particular, we notice the strongly polarized North pole (\cite{Schmid2011} observed the South pole) and the weak polarization in the higher latitude regions. A more quantitative analysis of our observations of Saturn and other solar system planets is currently in preparation.

\subsubsection{Henize 401}
The final class of objects observed during the first on-sky deployment of RITPIC were a number of reflection, planetary and proto-planetary nebulae. These objects are very challenging to observe due to their low surface brightness. On the other hand, nebulae rarely show the kind of steep intensity gradients that are seen in images of stars or solar system planets.  

\begin{figure*}[ht]
\centering
\includegraphics[width=\textwidth]{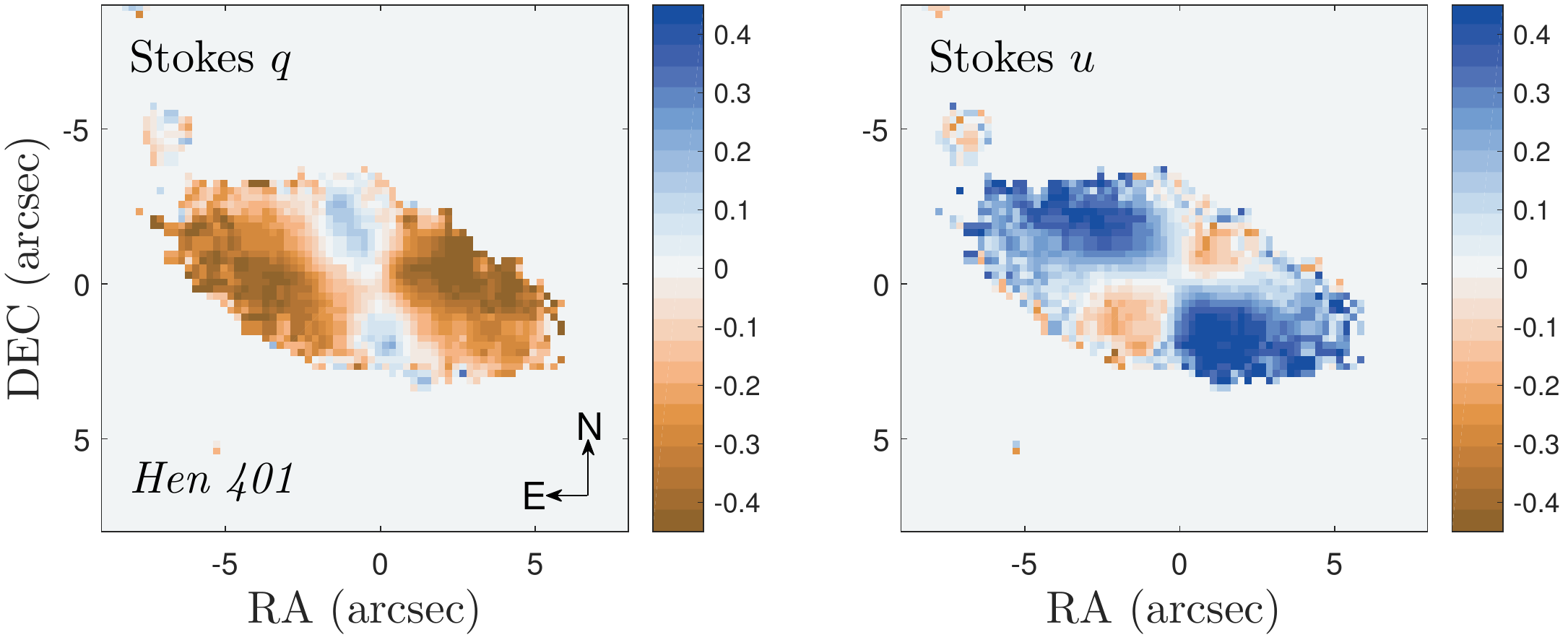}
\caption{Normalized Stokes parameters $q$ and $u$ calculated for Hen 401 using 5 300 second exposures in the R filter.} 
\label{fig:Hen401_stokes}
\end{figure*}

Henize 3-401 (IRAS 10178-5958) is a bipolar proto-planetary nebula, surrounding a Be post-AGB star. \textit{HST} imaging shows Hen 401 to be $\sim$20 $\times$ 5\arcsec. We acquired 5 300 s exposures in the R filter of Hen 401 on 2016, February 6. The normalized Stokes parameter maps for Hen 401 are shown in Figure \ref{fig:Hen401_stokes}. The nebula shows very strong polarization in both $q$ and $u$, reaching a maximum of $\sim$50\% in the lobes. The region near the central star is very weakly polarized. Additionally, these maps show spurious polarization detected for a star at the northeast tip of the nebula. This further exemplifies the challenges associated with the IFOV problem, and the need for the aperture photometry approach for unresolved sources.

\begin{figure*}[ht]
\centering
\includegraphics[width=\textwidth]{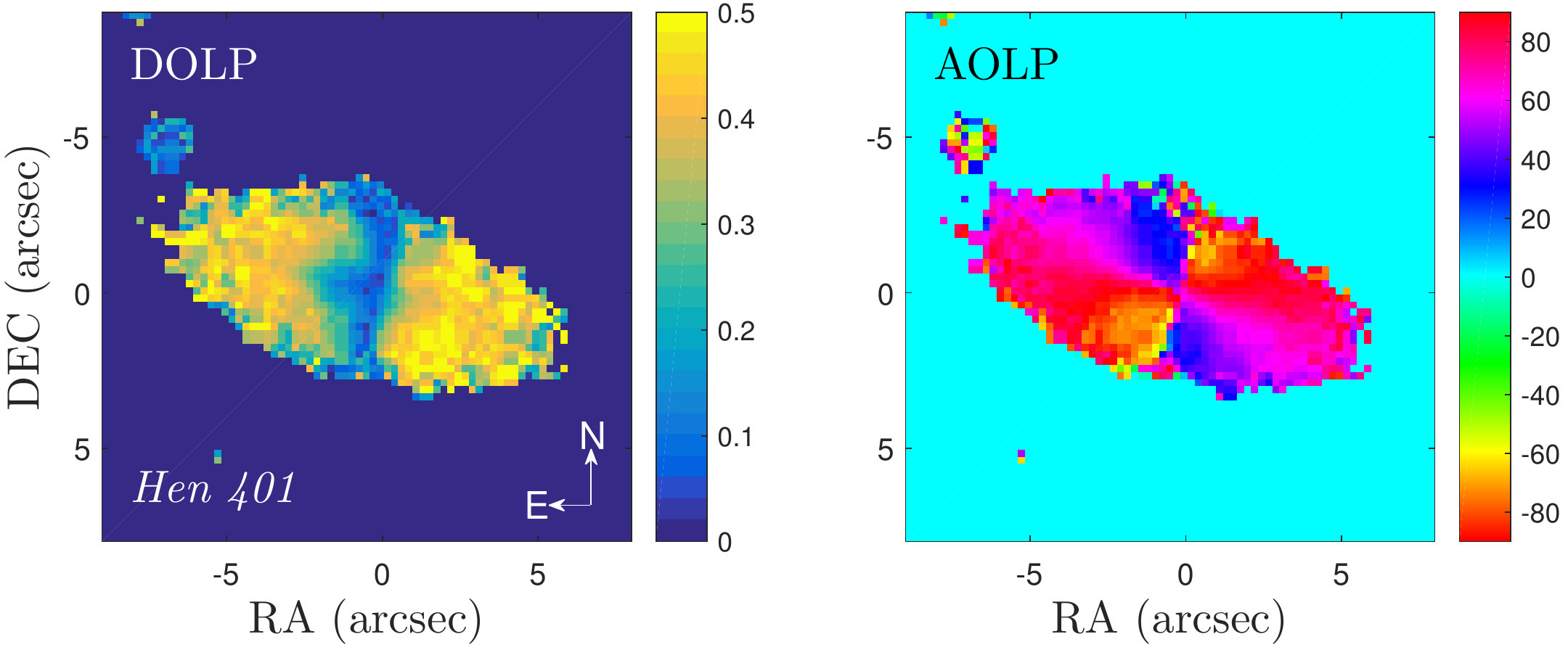}
\caption{The degree of linear polarization and angle of linear polarization maps for Hen 401, in the Bessel R band.} 
\label{fig:Hen401R_dolp}
\end{figure*}

The degree of linear polarization map for Hen 401 is shown in Figure \ref{fig:Hen401R_dolp}. The peak polarization in the lobes and the low polarization ``trench'' near the center agree with previous \textit{HST} polarimetry performed by \cite{Ueta2007}. This comparison is meant only as a ``sanity check''; a more rigorous comparison requires reprocessing of the raw HST data to simulate the effects of seeing-induced blurring and is the subject of future work. Nevertheless, the sharp transition near the central region is clearly resolved in RITPIC images, even with seeing-limited data. 

\section{Discussion of the Results}
\subsection{RITPIC's Calibration and Performance}
The current generation of polarization sensors, fabricated by aligning a micropolarizer array with an off-the-shelf imaging sensor, show a throughput of $\sim$30\% (for unpolarized light) and an average broadband contrast of $\sim$30:1. The performance is similar across the visible range (see Appendix \ref{sec:appendixMultiwavelengthCharacterization}). The throughput appears to be limited by the nonuniform fill-factor of the micropolarizer arrays, caused by opaque gaps around each micropolarizer. The contrast ratio is limited by several cross-talk mechanisms. Of these, the two most significant ones are diffraction of light by micropolarizers, which causes light transmitted by one micropolarizer to end up in a detector pixel designated for a neighboring polarizer and reflections within the glass substrate, as well as between the micropolarizer array and the detector surface.

In this work, we describe the design and performance of the RIT Polarization Imaging Camera - an imaging polarimeter based on a micropolarizer array. Polarimeters that only use the division-of-focal plane modulation scheme lack the precision afforded by the dual-beam dual-difference technique. Because the polarization is measured using intensities recorded by several different pixels, the performance of the polarimeter is ultimately limited by the ability to accurately characterize the pixel-to-pixel variations in the device. In this sense, the situation is very similar to that of conventional aperture photometry.  

Using the current characterization of RITPIC, we measured the very weakly polarized star HD 90156 to have a degree of polarization consistent with zero in the V and R bands, given the uncertainty in our measurements. Measurements in the B band appear to show some instrumental polarization in Stokes $u$. For the polarized star Vela X-1, our measurements are consistent with previous observations, within the uncertainty of $\sim0.3\%$ (see Table \ref{tbl:calibrationStandards}). 
In imaging mode, we are able to detect features that are polarized at the level of $\sim0.3\%$, as in the case of Saturn's disk.

\section{Conclusions and Future Outlook}

In this work, we presented the design, characterization and calibration of RITPIC, an imaging polarimeter based on a micropolarizer array. The performance of a polarimeter is challenging to specify, because it also depends on the source properties. The scatter in the estimation of the Stokes parameters derived from multiple measurements of the same source made with RITPIC (using $\sim$60\% of the CCD pixel well depth) is $\sim$0.2 - 1\%, depending on the seeing conditions. This means, using the current calibration, RITPIC cannot easily determine the polarization of sources with fractional polarization less than $\sim$0.5\%. However, sources with higher intrinsic polarization can be measured with sufficient confidence. For example, RITPIC's measurement of the polarization of Vela X-1 agrees with previous measurements at the 0.3\% level (see Table \ref{tbl:calibrationStandards}). Similarly, imaging polarimetry of Saturn shows structure at $\sim 0.3\%$ level. 

The quality of the polarimetry attainable with devices like RITPIC is ultimately limited by the quality of the photometric calibration, which relies on accurate characterization of the device parameters like throughput, contrast, and pixel orientation. As such, the photometric precision of RITPIC is limited by the same challenges that plague CCD-based differential photometry. As a general rule, the precision of differential photometry is limited to the level of a few millimagnitudes \citep{Everett2001}. Therefore, we expect that micropolarizer-based sensors can be calibrated to result in a polarimetric sensitivity of 0.1\% - 0.2\%, which corresponds to a photometric precision of 1 mmag and 2 mmag, respectively. 

The level of performance we achieved is comparable to that seen with other single-beam systems, like NICMOS and ACS (see the discussions in \cite{Hines2000, Batcheldor2009} and \cite{Avila2017}). Modern dual-beam polarimeters (imaging and non-imaging) routinely operate with sensitivities a factor of 10-100 lower than RITPIC. Micropolarizer-based polarimeters are not suitable for applications that require sensitivity greater than $\sim$0.1\%. On the other hand, this still leaves many science cases that can be studied with these detectors and what they lack in raw sensitivity, they make up with their low cost, compactness, reliability and ease of use. We identify several promising targets in the next section.

\subsection{Promising Science Cases}
Micropolarizer-based polarimeters are extremely versatile detectors and are suited for a wide range of observations. Any object (extended or unresolved) that is sufficiently bright (for shot noise-limited observations), with an intrinsic polarization $\gtrsim$0.5\% can be measured with these sensors in a relatively straightforward way; a more rigorous characterization should allow observation of objects with polarization as small as $\sim$0.2\%. 

The inherent stability, compactness, and ease of use (as compared to conventional polarimeters) afforded by micropolarizer-based sensors makes them especially suited for deployment on smaller ground-based telescopes, as well as space-based and airborne platforms. We conclude this section with a brief outline of science targets that well appear suited for these polarimeters:

\textit{The Solar Corona.} The corona has been studied with polarimetric techniques for over a century. Broadband polarimetry allows an independent way to constrain the electron density in the corona, while spectropolarimetry opens a window into more detailed kinematics. These polarimeters allow routine imaging polarimetry of the corona, with higher precision than has been possible by past polarimeters, based on photographic film.

\textit{Solar System Planets.} The surfaces and atmospheres of planets in the Solar System show changes on scales that range from hours to years \citep{Gehrels1969} and polarization fractions that are easily accessible to instruments like RITPIC. High cadence, long term polarimetry of the inner planets and the gas giants may reveal interesting insights into the seasonal and daily aspects of their atmospheres. 

\textit{Comets.} Polarimetry is a key tool in the study of comets. Over the course of their orbit, comets allow measurements at a wide range of phase angles, which helps constrain the microscopic geometry of the scattering particles in the coma. Because these sensors are inexpensive, they can be used to develop a world-wide network of observers to provide continuous monitoring of comets as they enter and leave the inner Solar System. The stability of these polarimeters makes it relatively straightforward to compare measurements made with different telescopes. 

\textit{Proto-planetary Nebulae.}  Stars in the post-asymptotic giant branch stage of their evolution often show strong polarization and rapid variability. These objects are often accompanied by nebulae that also show complex and prominent polarization structure. Polarimeters like RITPIC on 1 m class telescopes provide an excellent platform for long-baseline monitoring of these objects.

Micropolarizer-based polarization sensors represent the first ``general purpose'' polarimeter. Although these devices are not the most precise polarimeters that have been built, they are the first that can be taken off a telescope and immediately used with a microscope, or any other imaging system. The calibration and data analysis techniques developed for these devices can be easily used to study the Earth's surface, fluorophores on a microscope slide, or the structure of human skin. This flexibility may end up being the most useful property of these devices. 

\acknowledgments

The authors acknowledge the support and funding provided by Moxtek, Inc. In particular, we thank Ray West and Roger Ketcheson. DV would like to thank Todd Henry and the staff at CTIO, especially Hernan Tirado and Humberto Orrego. This work has benefitted from many insightful discussions with Michael Richmond and Dean Hines. The multi-wavelength characterization data was acquired by Lee Bernard, while he was an REU student in RIT's Center for Imaging Science.

\appendix
\section{Multi-wavelength Characterization of RITPIC}
\label{sec:appendixMultiwavelengthCharacterization}
\begin{figure*}[ht]
\centering
\includegraphics[width=\textwidth]{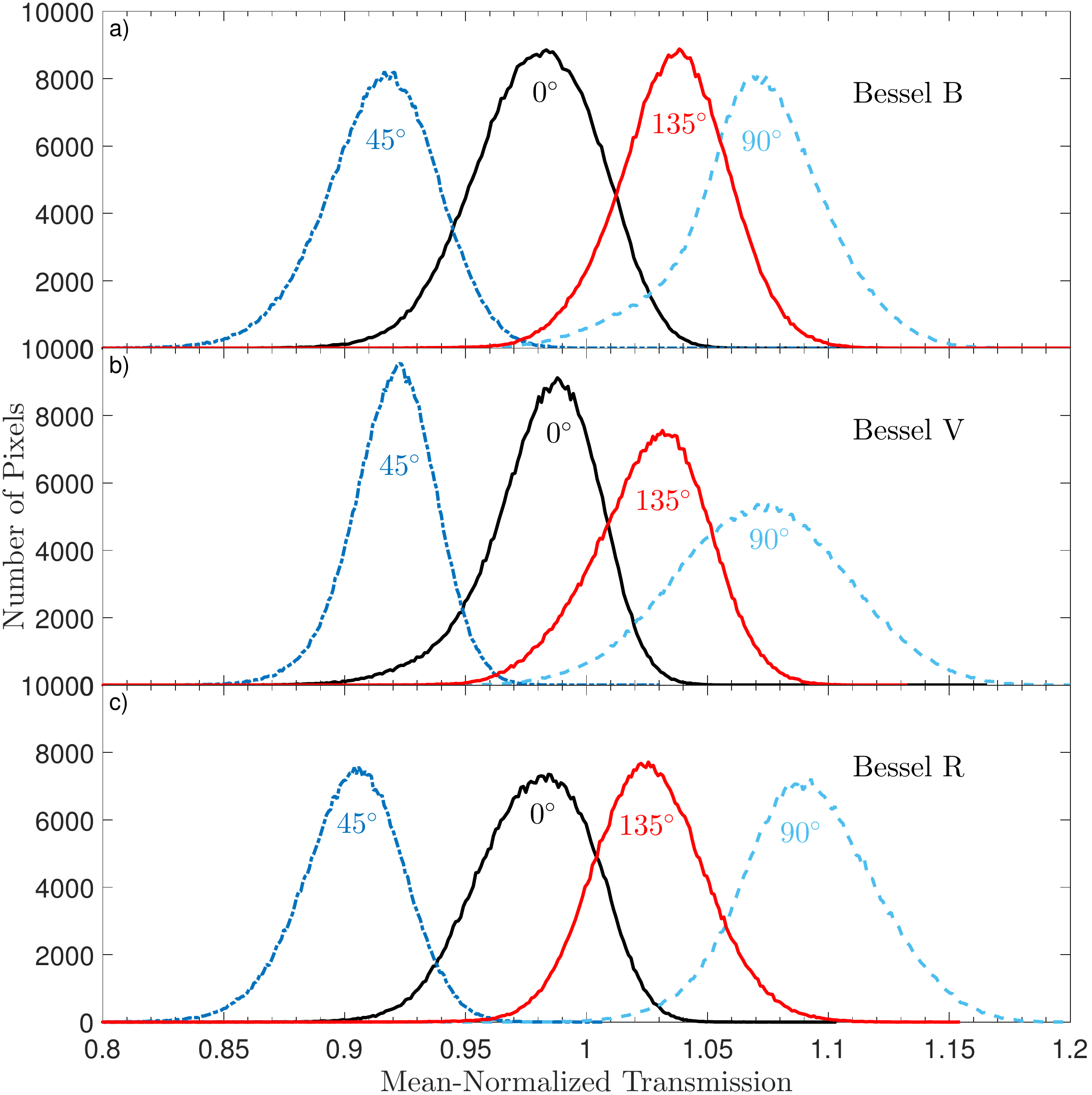}
\caption{The relative throughput of RITPIC is similar across the visible range. The device response is most uniform in the V band.}
\label{fig:throughput_BVR}
\end{figure*}

\begin{figure*}[ht]
\centering
\includegraphics[width=\textwidth]{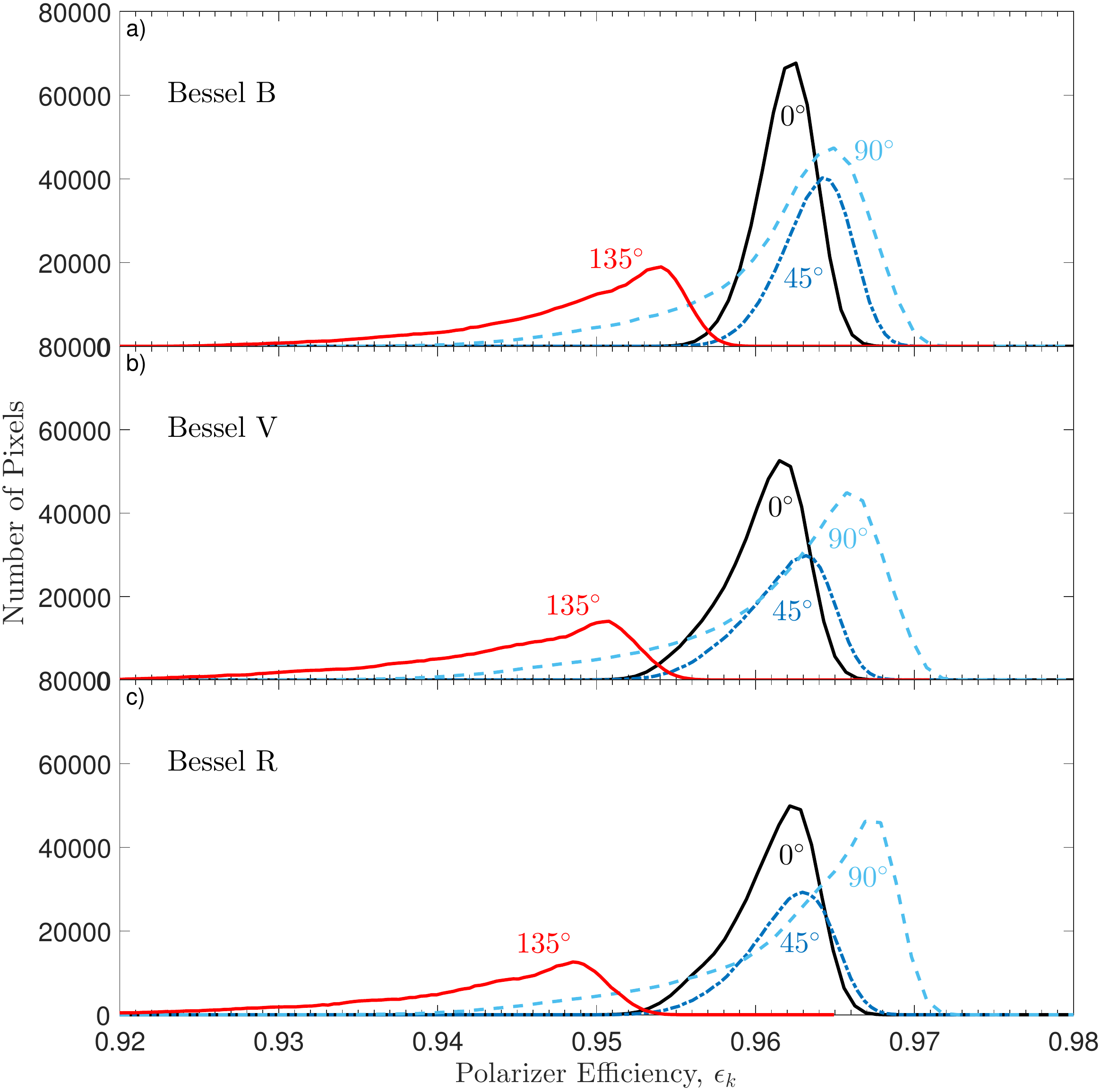}
\caption{The polarizer efficiency of RITPIC's pixels does not change significantly across the visible range. Nevertheless, the differences are large enough to require different calibrations for each passband.}
\label{fig:efficiency_BVR}
\end{figure*}

Polarimeters based on wire grid micropolarizer arrays operate over a large spectral range. Nevertheless, the instrumental response varies with wavelength. In this section we show the relative throughput (Figure \ref{fig:throughput_BVR} characterization and the polarizer efficiency (Figure \ref{fig:efficiency_BVR}) for the Bessel BVR filters.

\begin{figure}[ht]
\centering
\includegraphics[width=0.5\textwidth]{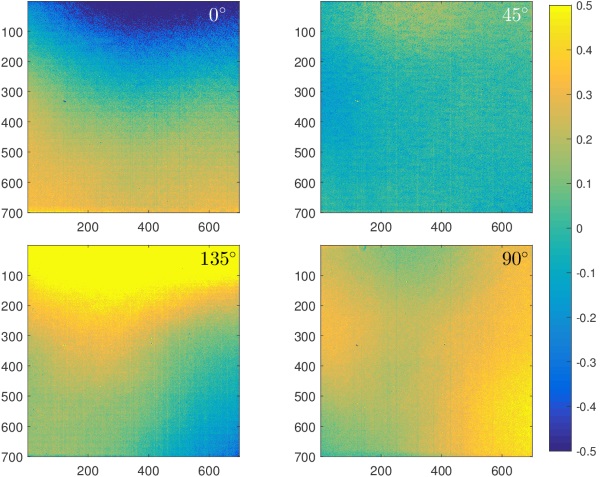}
\caption{This composite image shows the deviation from the nominal orientation for each pixel orientation. For example, all $0^{\circ}$ pixels are quite close to their nominal orientation, whereas the $135^{\circ}$ degree pixels are oriented slightly less than $135^{\circ}$ near the top of the array and slightly more near the bottom.}
\label{fig:spatial_pk}
\end{figure}

Furthermore, the orientation of each pixel shows variation on large scales across the detector (Figure \ref{fig:spatial_pk}, similar to the throughput differences shown in Figure \ref{fig:spatialDistribution_tk}. This is another manifestation of the ``effective orientation'' that we measure during the characterization process (see Section \ref{sec:performanceCharacterization}). 

\section{Analysis of Uncertainty in Aperture Polarimetry}
\label{sec:appendixAnalysisOfUncertainty}
In general, the uncertainty of a polarimetric measurement depends on the absolute polarization of the source, the intrinsic variability of the source, and the noise associated with the measurement \citep{Simmons1985} (due to the combined effects of mechanisms described in Sec. \ref{sec:noiseSources}). In our analysis of uncertainty, we consider four different mechanisms:
\begin{enumerate}
    \item Uncertainty due to Poisson noise (with contributions from read noise and bias+dark subtraction), \textit{i.e.} $\sqrt{N_e}/N_e$.
    \item The uncertainty associated with the aperture photometry in each subframe.
    \item The goodness of the model fit associated with the demodulation process, estimated using analysis of residuals (described below).
    \item The overall scatter of measurements acquired with a series of exposures.
\end{enumerate}

\begin{figure}[ht]
    \centering
    \includegraphics[width=0.5\textwidth]{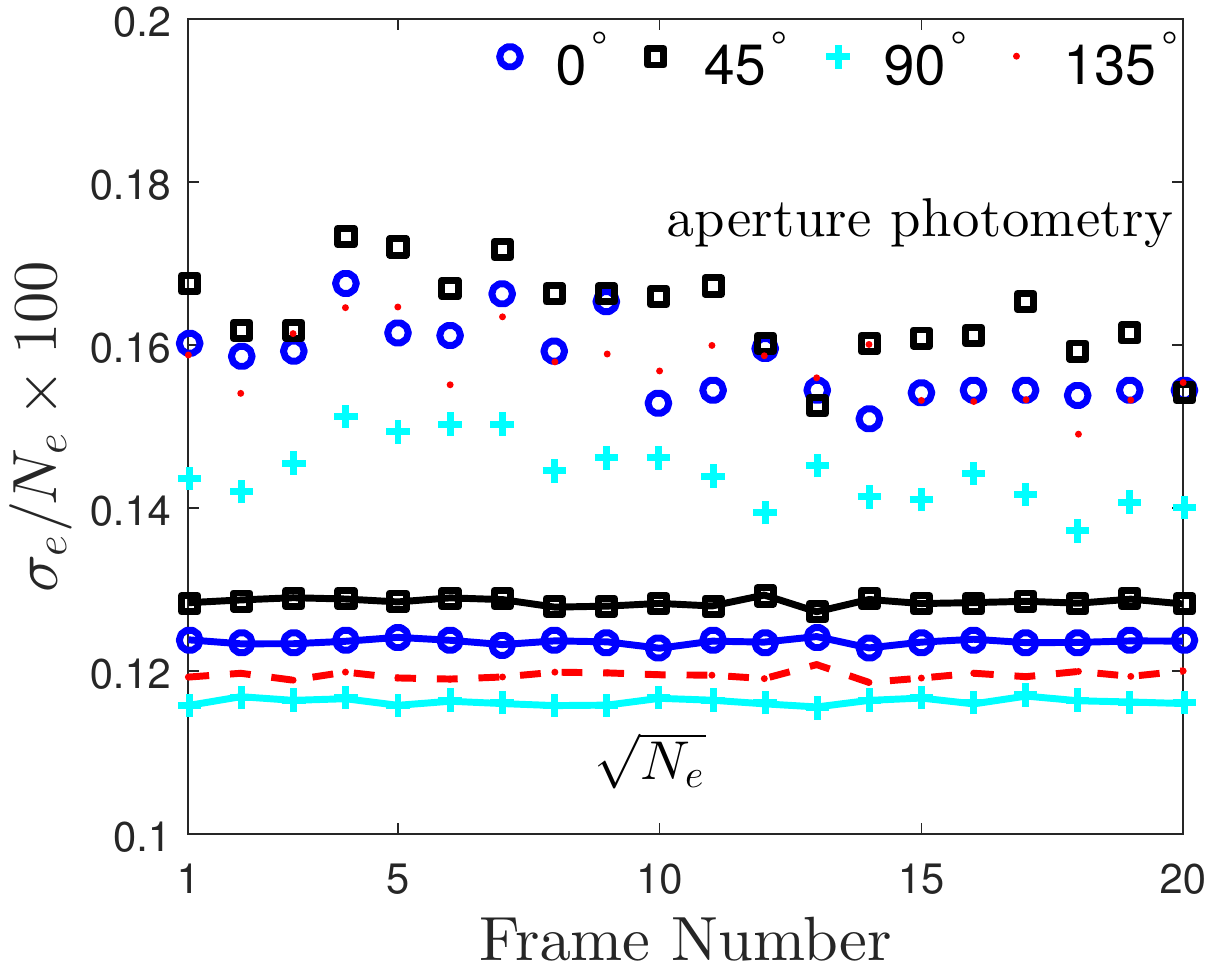}
    \caption{Photometric errors in a set of 20 observations of HD 90156 in the Bessel R band, associated with the flux estimation in the four subframes. The bottom plots show a shot noise estimate using only the total flux in each subframe ($\sqrt{N_e}$), whereas the top plots show a more realistic estimate, associated with the aperture photometry process.} 
    \label{fig:hd90156_photometricErrors}
\end{figure}

Estimates of the photometric SNR, which is mostly due to Poisson statistics, are shown in Figure \ref{fig:hd90156_photometricErrors}. Here we show the estimated SNR using only the fluxes estimated in each subframe, $\sqrt{N_e}$, and the SNR after background subtraction. As one might expect, the overall SNR after aperture photometry is slightly worse than that due to pure shot noise. Systematic differences in pixel throughput reveal themselves as systematically different SNRs between each pixel orientation. Furthermore, scatter due to variations in seeing can be seen in the top set of plots. 

\begin{figure}[ht]
    \centering
    \includegraphics[width=0.5\textwidth]{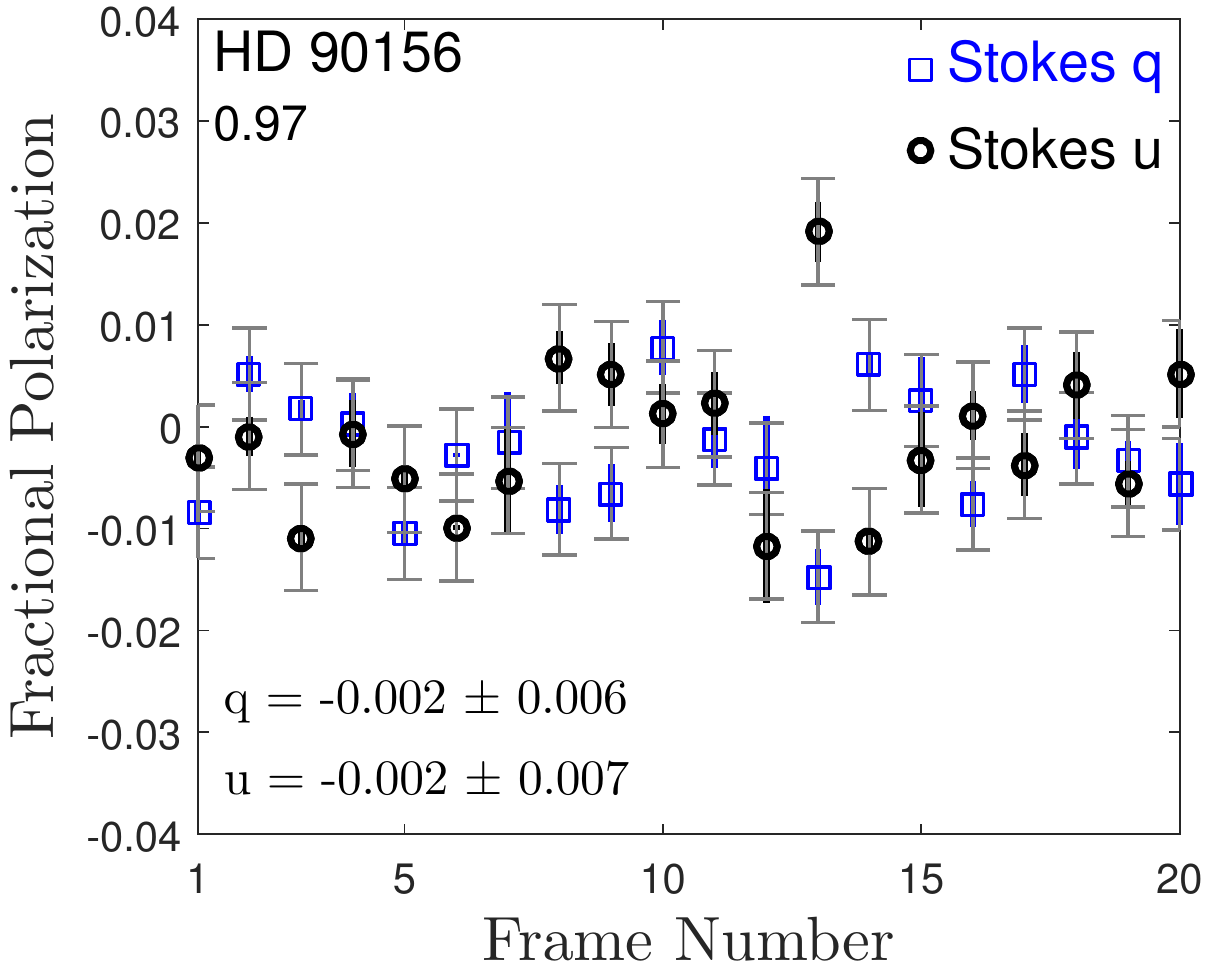}
    \caption{Estimation of the normalized Stokes parameters, $q$ and $u$, for HD 90156 using a set of 20 observations in the Bessel R band. The median and standard deviation of the 20 measurements are $q = -0.003\pm0.006$ and $u=0.002\pm0.008$. The errors bars with horizontal marks represent the expected error due to the scatter in the flux estimation and the purely vertical bars show the residuals from the model fit.} 
    \label{fig:hd90156_polarimetricErrors}
\end{figure}

We estimate the Stokes parameters using a linear least squares fit to the fluxes obtained in each subframe using aperture photometry; in practice this is done using a matrix pseudo-inversion (Figure \ref{fig:hd90156_polarimetricErrors}). We estimate the uncertainty of the Stokes parameter calculation using three techniques. First, we use formal (linear) error propagation and the photometric uncertainty associated with aperture photometry. A complimentary approach involves the analysis of residuals of the least squares fit used to estimate the Stokes parameters. The variance associated with each measurement, $\sigma^2$, is calculated using the unbiased estimator $\hat{\sigma}^2$ \citep{Bajorski2011} as follows,

\begin{equation}
\label{eq:varianceEstimator}
\hat{\sigma}^2=\frac{1}{n-k-1}\sum_{i=1}^{n}(y_i - \hat{y}_i)^2  = \sum_{i=1}^{n}(y_i - \hat{y}_i)^2,
\end{equation}

where $y_i$ is an observed intensity and $\hat{y}_i$ is the model fit. The number of measurements is $n=4$ and the number of regression coefficients is $k+1=3$. These residuals are themselves subject to photometric noise, and their mean is roughly comparable to the uncertainty estimated through formal propagation of photometric errors. 

The last, and most straightforward, approach to estimate the uncertainty of the Stokes parameters is using the scatter of the 20 independent measurements shown in Figure \ref{fig:hd90156_polarimetricErrors}. In this case, the standard deviation for Stokes $q$ and $u$ is $\sigma_q=0.006$ and $\sigma_u=0.007$. This scatter is larger than the error bars on a single measurement, due to photometric uncertainty. However, it is roughly consistent with the frame-to-frame scatter seen in flux estimation obtained by aperture photometry. In general, the frame-to-frame scatter in the individual flux estimates appears to explain most of the scatter seen in the Stokes parameter estimation. One clear outlier is frame number 13, where the estimated Stokes $u$ is unusually high; a manual inspection of this frame did not reveal any obvious mechanism for this discrepancy. 

\section{Common Errors in MPA-based Imaging Polarimeters}
\label{sec:appendixImagingPolarimetry}
To better illustrate the kinds of systematic effects that appear due to improper calibration and data analysis, the next few figures show the estimation of Stokes $q$ and $u$ and how it changes as the data analysis process is improved.  

\begin{figure*}[ht]
\centering
\includegraphics[width=\textwidth]{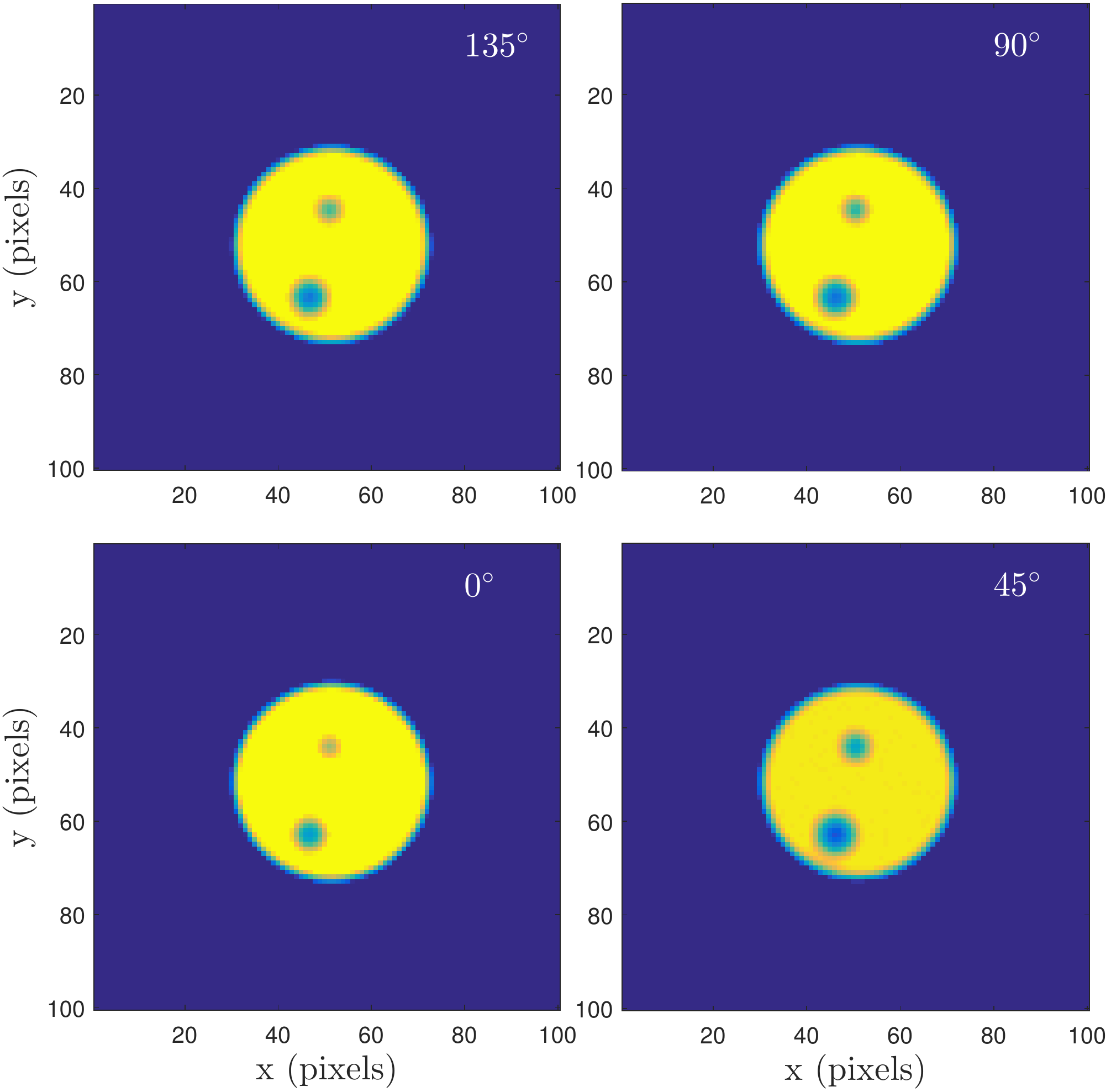}
\caption{The subframes corresponding the four pixel orientations of a synthetically generated image, with Poisson noise and a photometric SNR of $\sim$100. The two darker spots in the object are polarized, which is why they appear larger or smaller, depending on the subframe. The object is shifted $\pm0.5$ pixels in each frame, with respect to the other frames.}
\label{fig:imagingSubframes}
\end{figure*}

\begin{figure*}[ht]
\centering
\includegraphics[width=\textwidth]{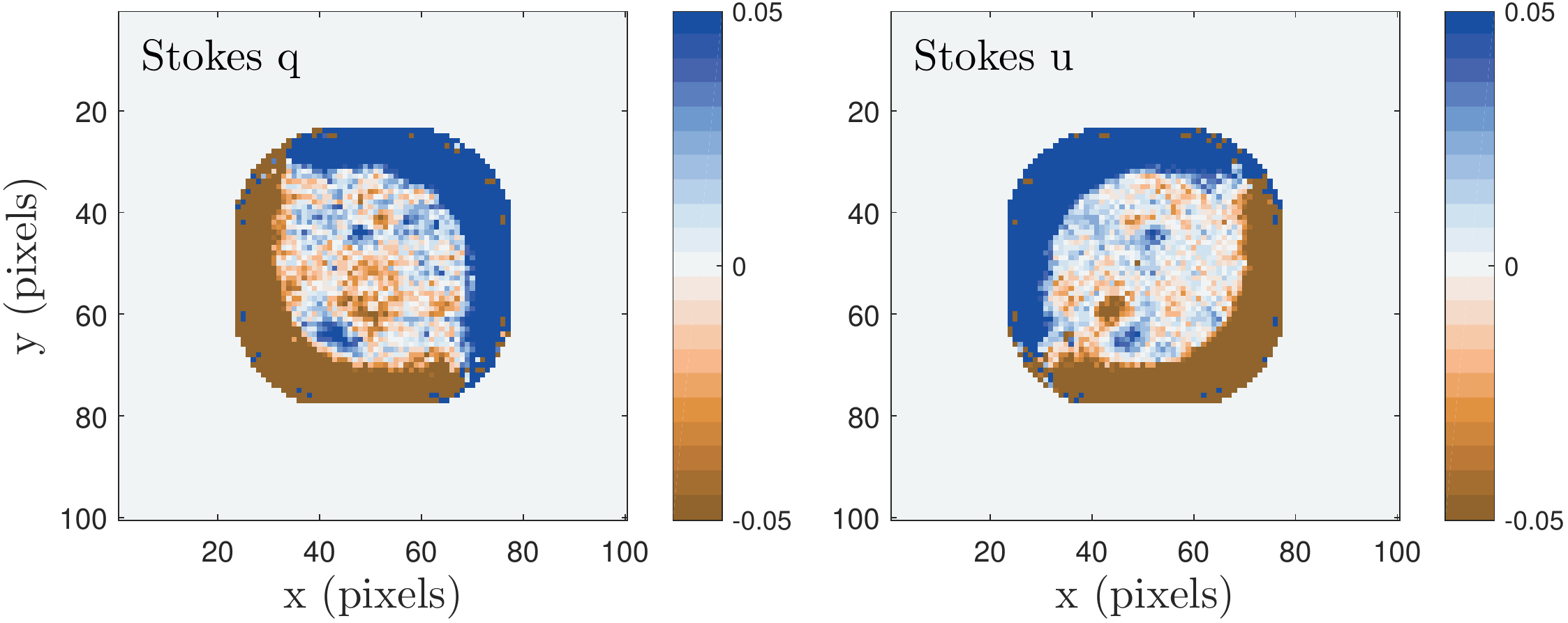}
\caption{The estimation of Stokes $q$ and $u$ parameters for the object in Figure \ref{fig:imagingSubframes}, without proper calibration. The two smaller spots have intrinsic values $q$ = 0.01 and $u$ = 0.}
\label{fig:imagingNoCalibration}
\end{figure*}

To illustrate the data analysis process, we generated a synthetic planet-like object with no polarization across its face, except two weak circular features. The raw, synthetic counts are shown in Figure \ref{fig:imagingSubframes}. The original image has a mean shot noise SNR per pixel of $\sim$150. The two smaller circles are polarized with fractional polarization $p$ = 0.01 and $\psi$ = $0^{\circ}$, resulting in $q$ = 0.01 and $u$ = 0. First, let's look at the resulting $q$ and $u$ maps without flat field correction or image registration (Figure \ref{fig:imagingNoCalibration}). 

\begin{figure*}[ht]
\centering
\includegraphics[width=\textwidth]{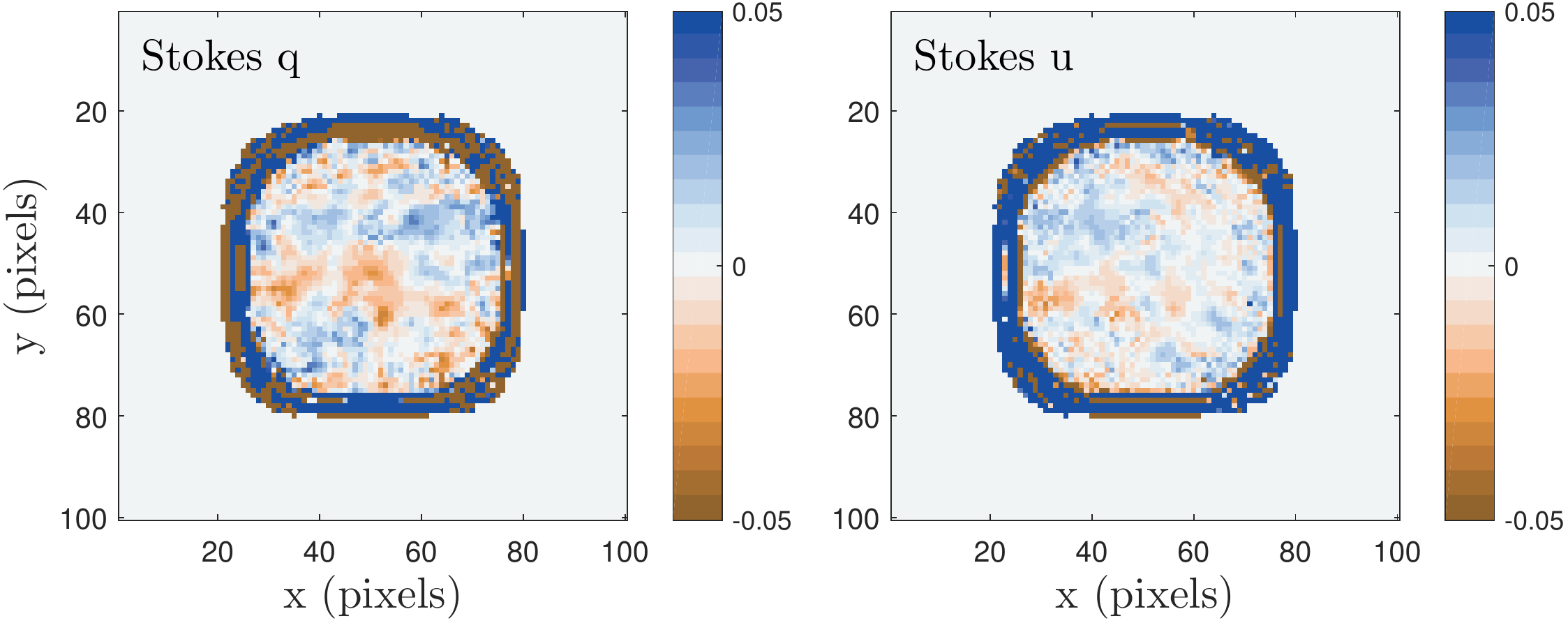}
\caption{The estimation of Stokes $q$ and $u$ parameters for the object in Figure \ref{fig:imagingSubframes}, with image registration but no flat field correction.}
\label{fig:imagingNoFlats}
\end{figure*}

The Stokes $q$ and $u$ maps in Figure \ref{fig:imagingNoCalibration} show several characteristic errors associated with micropolarizer-based polarimeters. Most of the image has zero polarization, however the estimation shows regions of very high polarization near the edges of the object. In these areas, the steep gradients around the edges and lower photometric SNR are causing large systematic errors; the two smaller polarized spots show a bi-lobed structure in the polarization maps for the same reason. These large features are eliminated when the subframes are properly registered (Figure \ref{fig:imagingNoFlats}). 

\begin{figure*}[ht]
\centering
\includegraphics[width=\textwidth]{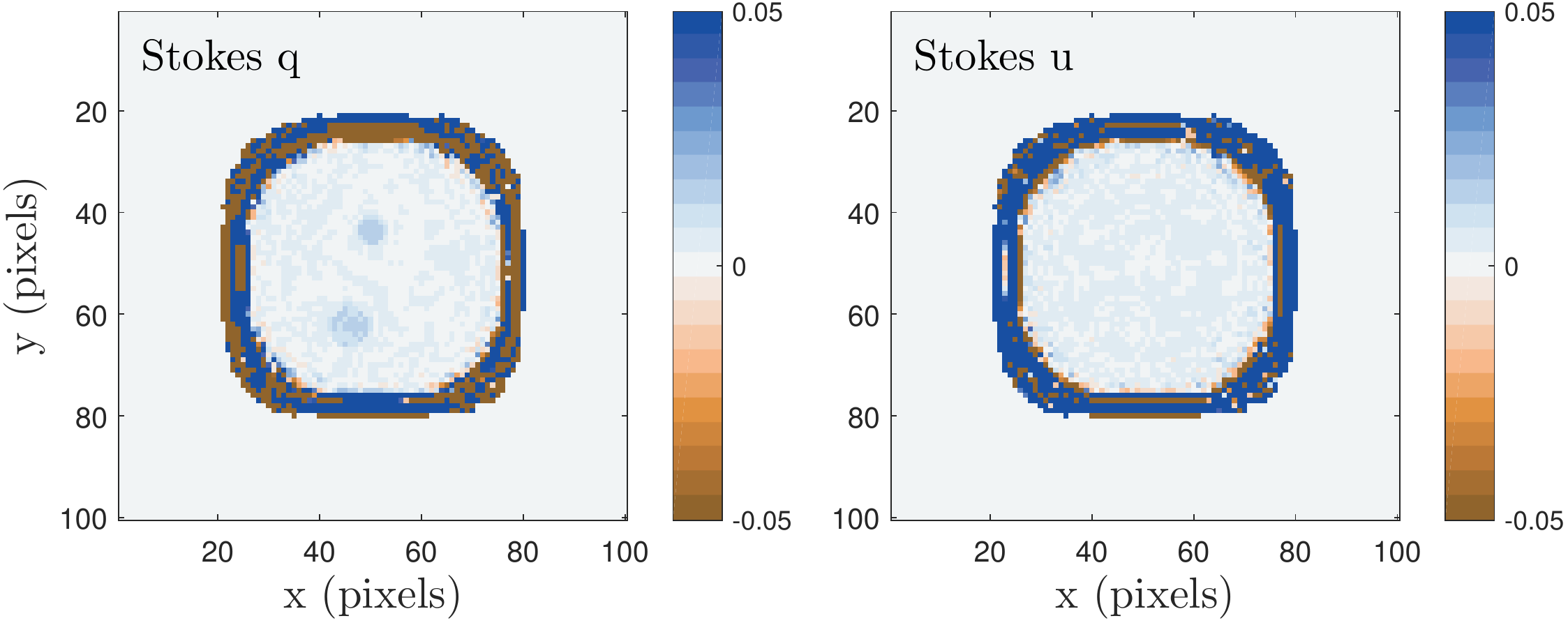}
\caption{The estimation of Stokes $q$ and $u$ parameters for the object in Figure \ref{fig:imagingSubframes}, with image registration and flat field correction.}
\label{fig:imagingWithNoise}
\end{figure*}

\begin{figure*}[ht]
\centering
\includegraphics[width=\textwidth]{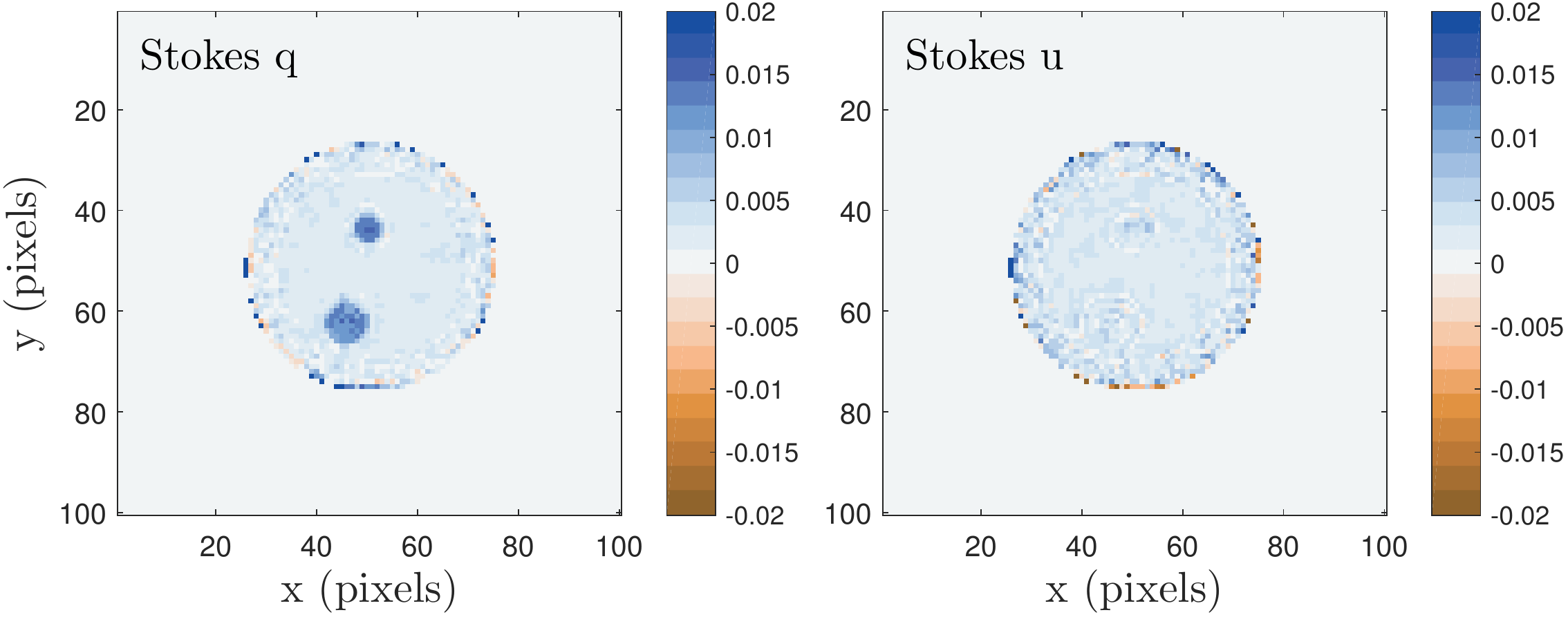}
\caption{The estimation of Stokes $q$ and $u$ parameters for the object in Figure \ref{fig:imagingSubframes}, with image registration and flat field correction.}
\label{fig:imagingWithoutNoise}
\end{figure*}

When the subframes are properly registered, the the bi-lobed asymmetric features disappear, but structure at $\sim$2\% level is seen across the face of the object. This is due to pixel-to-pixel variations in throughput that are not corrected. The next step is to correct the throughput variations using flat fields. The resulting polarization maps are shown in Figure \ref{fig:imagingWithNoise}.
Performing the throughput correction removes the several-percent errors seen across the face of the object, revealing the weakly polarized spots. The remaining artifacts around the edge of the object are due to low per-pixel SNR. Masking off pixels with SNR less than 30 removes most of the erroneous polarization (Figure \ref{fig:imagingWithoutNoise}).

\newpage
\clearpage

\bibliographystyle{aasjournal}
\bibliography{Bibliography}{}



\end{document}